\documentclass[manuscript,screen]{acmart}
\usepackage{graphicx}
\usepackage{cleveref}
\usepackage{multirow, xurl, booktabs}
\usepackage{soul}
\usepackage{subfigure}
\usepackage{xcolor}
\definecolor{revision}{rgb}{0.0, 0.5, 1.0}

\title{GUing: A Mobile GUI Search Engine using a Vision-Language Model}

\author{Jialiang Wei}
\email{jialiang.wei@mines-ales.fr}
\orcid{0009-0008-6028-1576}
\affiliation{
  \institution{EuroMov Digital Health in Motion, Univ Montpellier, IMT Mines Ales}
  \city{Ales}
  \country{France}
}

\author{Anne-Lise Courbis}
\email{anne-lise.courbis@mines-ales.fr}
\orcid{0000-0002-7530-4661}
\affiliation{
  \institution{EuroMov Digital Health in Motion, Univ Montpellier, IMT Mines Ales}
  \city{Ales}
  \country{France}
}

\author{Thomas Lambolais}
\email{thomas.lambolais@mines-ales.fr}
\orcid{0000-0002-4462-1277}
\affiliation{
  \institution{EuroMov Digital Health in Motion, Univ Montpellier, IMT Mines Ales}
  \city{Ales}
  \country{France}
}

\author{Binbin Xu}
\email{binbin.xu@mines-ales.fr}
\orcid{0000-0002-0822-5250}
\affiliation{
  \institution{EuroMov Digital Health in Motion, Univ Montpellier, IMT Mines Ales}
  \city{Ales}
  \country{France}
}

\author{Pierre Louis Bernard}
\email{pierre-louis.bernard@umontpellier.fr}
\orcid{0000-0002-0023-4531}
\affiliation{
  \institution{EuroMov Digital Health in Motion, Univ Montpellier, IMT Mines Ales}
  \city{Montpellier}
  \country{France}
}

\author{Gérard Dray}
\email{gerard.dray@mines-ales.fr}
\orcid{0000-0003-1525-5682}
\affiliation{
  \institution{EuroMov Digital Health in Motion, Univ Montpellier, IMT Mines Ales}
  \city{Ales}
  \country{France}
}

\author{Walid Maalej}
\email{walid.maalej@uni-hamburg.de}
\orcid{0000-0002-6899-4393}
\affiliation{
  \institution{University of Hamburg}
  \city{Hamburg}
  \country{Germany}
}

\begin{document}

\begin{abstract}
Graphical User Interfaces (GUIs) are central to app development projects. 
App developers may use the GUIs of other apps as a means of requirements refinement and rapid prototyping or as a source of inspiration for designing and improving their own apps.
Recent research has thus suggested retrieving relevant GUI designs that match a certain text query from screenshot datasets acquired through crowdsourced or automated exploration of GUIs.
However, such text-to-GUI retrieval approaches only leverage the textual information of the GUI elements, neglecting visual information such as icons or background images. 
In addition, retrieved screenshots are not steered by app developers and lack app features that require particular input data.

To overcome these limitations, this paper proposes GUing, a GUI search engine based on a vision-language model called GUIClip, which we trained specifically for the problem of designing app GUIs. 
For this, we first collected from Google Play app introduction images which display the most representative screenshots and are often captioned (i.e.~labelled) by app vendors.
Then, we developed an automated pipeline to classify, crop, and extract the captions from these images.
This resulted in a large dataset which we share with this paper: including 303k app screenshots, out of which 135k have captions.
We used this dataset to train a novel vision-language model, which is, to the best of our knowledge, the first of its kind for GUI retrieval. 
We evaluated our approach on various datasets from related work and in a manual experiment. 
The results demonstrate that our model outperforms previous approaches in text-to-GUI retrieval achieving a Recall@10 of up to 0.69 and a HIT@10 of 0.91.
We also explored the performance of GUIClip for other GUI tasks including GUI classification and sketch-to-GUI retrieval with encouraging results.
\end{abstract}




\maketitle

\section{Introduction}
The Graphical User Interfaces (GUI) of mobile apps serve as the primary means of interaction between users and their devices.
A well-designed GUI streamlines the navigation process, facilitates the accomplishment of user tasks, and improves the overall user experience --- contributing towards a higher user engagement and retention \cite{Hassan:StudyingBadUpdates:2020,Chen:HowShouldImprove:2021}.
Furthermore, a modern, attractive, and user-friendly GUI can potentially differentiate an app from its market counterparts, amplifying its likelihood of success in the highly competitive app market \cite{Moran:AutomatedReportingGUI:2018, Martens:ReleaseEarlyRelease:2019}.
To design a good GUI, developers often create multiple prototypes with varying fidelity: from low-fidelity sketches for brainstorming to high-fidelity GUIs for testing and optimisation. 
GUI prototypes are used in interviews or workshops to discuss, refine, and validate requirements, helping reduce misunderstanding and ultimately saving resources.

In this context, app developers often explore other GUIs of related apps as a way for rapid prototyping and a source of inspiration to design and improve their own apps.
Numerous approaches have thus been suggested to retrieve relevant GUIs from existing apps. 
Researchers have proposed GUI retrieval approaches that accept sketches \cite{Huang:SwireSketchbasedUser:2019,Mohian:PSDoodleFastApp:2022,Mohian:SearchingMobileApp:2023}, wireframes \cite{Deka:RicoMobileApp:2017,Liu:LearningDesignSemantics:2018,Chen:WireframebasedUIDesign:2020}, or screenshots \cite{Li:Screen2vecSemanticEmbedding:2021,Bunian:VINSVisualSearch:2021} as input to locate similar or fitting designs.
While certainly useful, these approaches require a preliminary graphical prototype, which might not be available or might be too restrictive in early ideation phases.
Therefore, when only general ideas and textual requirements are available, text-to-GUI retrieval approaches would be particularly useful. 

\begin{figure*}[!htb]
    \centerline{\includegraphics[width=0.8\textwidth]{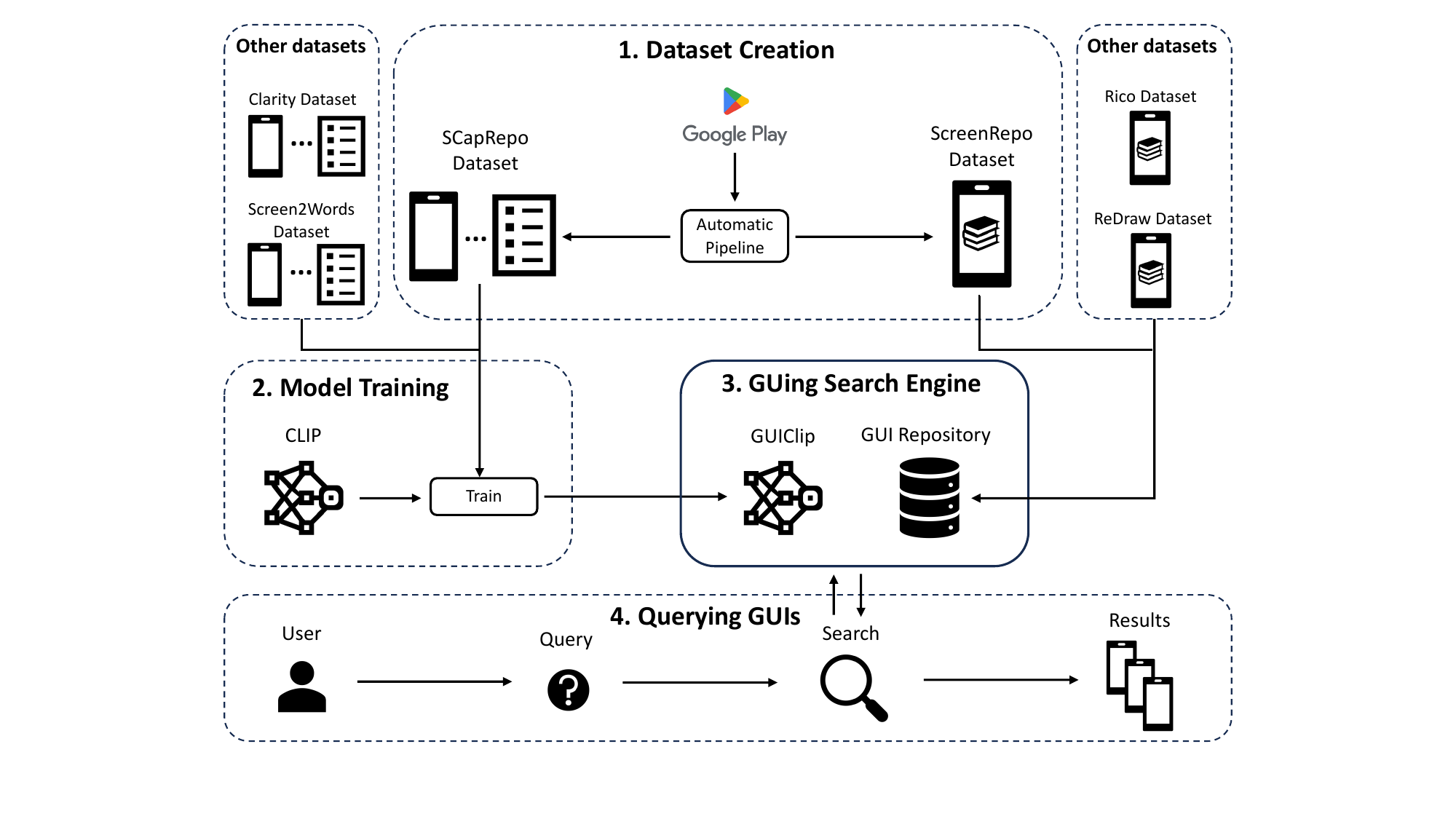}}
    \caption{Overview of our approach: including the creation of the datasets, the training of the vision-language model GUIClip, the development of the GUI search engine, and the process of querying on the engine.}
    \Description{Overview of our approach}
    \label{fig:overview}
\end{figure*}

Recent GUI search engines, such as GUIGLE \cite{Bernal-Cardenas:GuigleGUISearch:2019} and RaWi \cite{Kolthoff:DatadrivenPrototypingNaturallanguagebased:2023}, enable their users to search GUI datasets using text queries. 
Both engines use app metadata and GUI-related text for query matching. 
GUIGLE adopts basic keyword matching, while RaWi calculates semantic similarity based on a BERT model \cite{Devlin:BERTPretrainingDeep:2019}. 
However, these text-based retrieval approaches solely use textual information available within the GUI such as text blocks or button labels, neglecting key visual information such as images, layouts, and backgrounds. 
Furthermore, the underlying GUI datasets (namely Rico \cite{Deka:RicoMobileApp:2017} and ReDraw \cite{Moran:MachineLearningBasedPrototyping:2020}) are created by crowdsourced or automated exploration of app screens at runtime. 
However, the access to certain screens may require authorisation or initial configurations, which is often skipped in automated exploration.
Even with a crowdsourced exploration, some app features may not be captured, particularly those requiring substantial or specific input data, such as dashboard pages with charts.
Thus, screenshots crawled at runtime may fail to comprehensively capture essential features of the app.

In this paper, we propose a novel GUI search engine based on a vision-language model trained with app introduction images from Google Play as shown on Figure \ref{fig:overview}.
Recent vision-language models, such as CLIP \cite{Radford:LearningTransferableVisual:2021}, BLIP \cite{Li:BLIPBootstrappingLanguageImage:2022}, and BLIP-2 \cite{Li:BLIP2BootstrappingLanguageImage:2023}, are trained on large-scale image-caption data using contrastive learning.
These models have the ability to transform images and text into a \textit{multimodal} embedding, ensuring that semantically similar images and texts are mapped closely in the embedding space. 
Thus, by computing the similarity between the images and the textual query, these models can be used for text-to-image retrieval tasks.
As prerequisite a large screenshot-caption dataset is needed. 
Since, currently available datasets (Screen2Words \cite{Wang:Screen2WordsAutomaticMobile:2021} and Clarity \cite{Moran:EmpiricalInvestigationUse:2022}) are inadequate for training a vision-language model (as our evaluation shows in Section \ref{sec:eval-result}), we have created a new large screenshot-caption dataset.

\begin{figure*}[!htb]
    \subfigure[Surrounded Screenshots]{\includegraphics[width=0.3\textwidth]{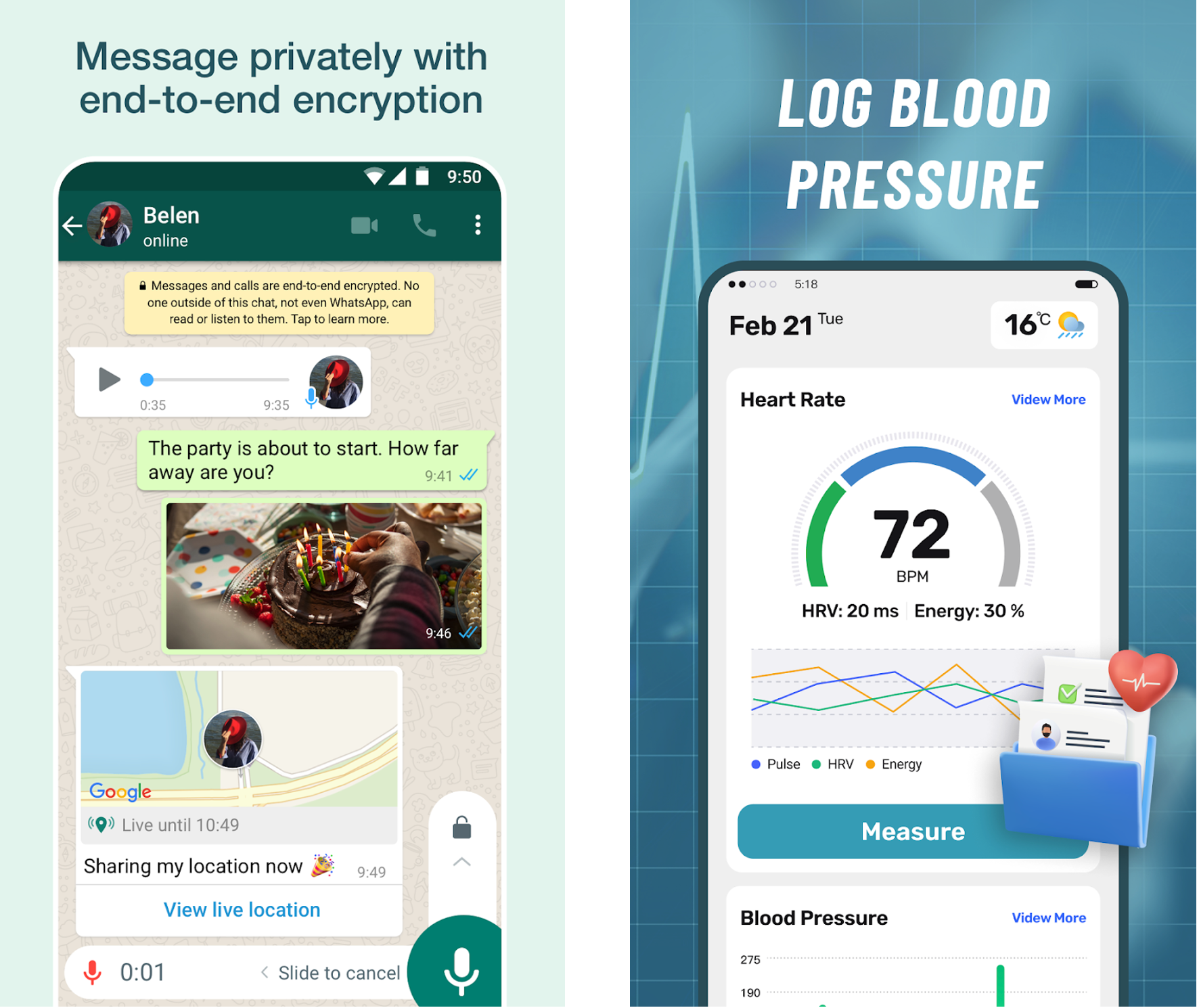}} 
    \hspace{1.5em}
    \subfigure[Screenshots]{\includegraphics[width=0.3\textwidth]{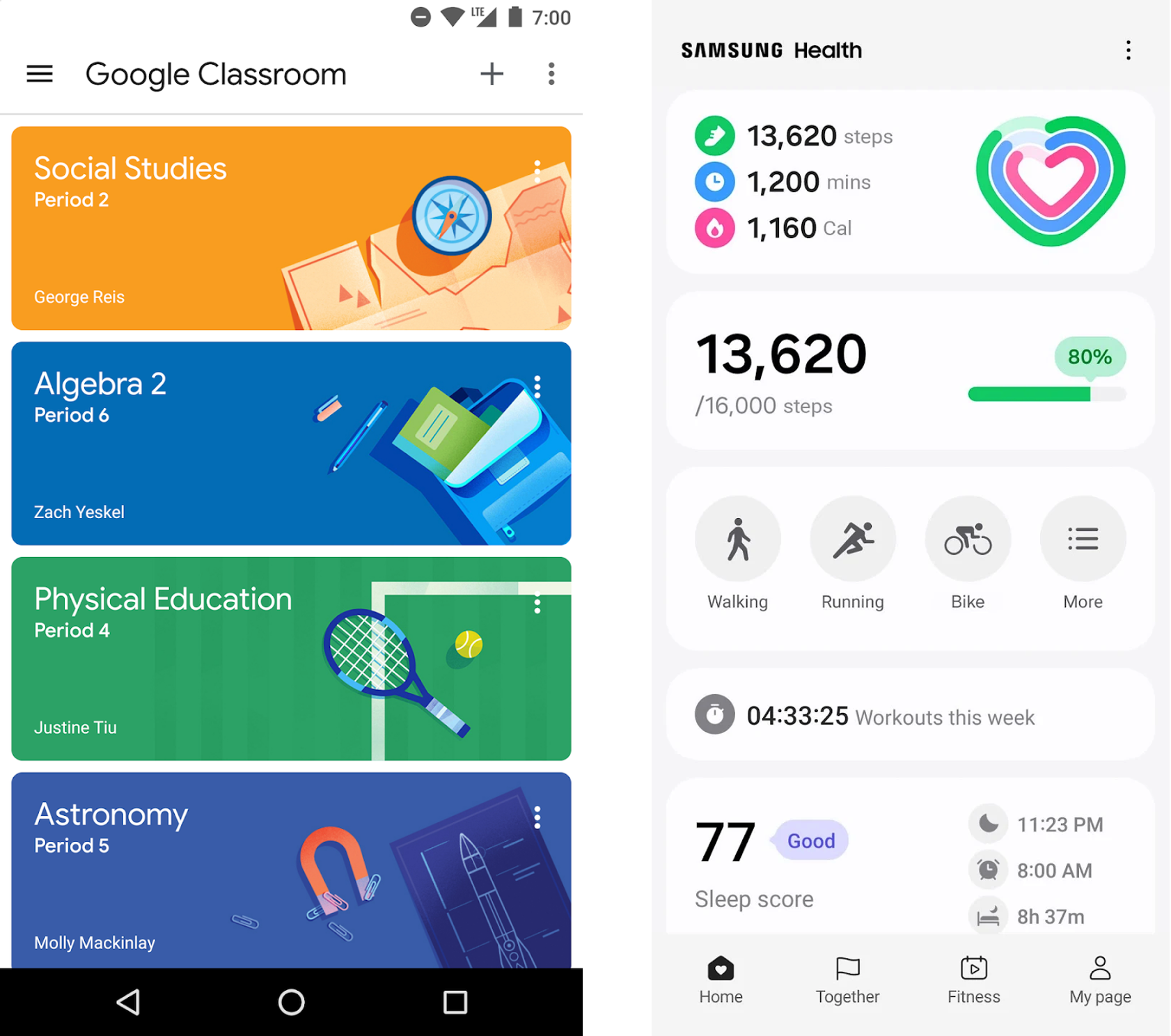}} 
    \hspace{1.5em}
    \subfigure[Irrelevant]{\includegraphics[width=0.3\textwidth]{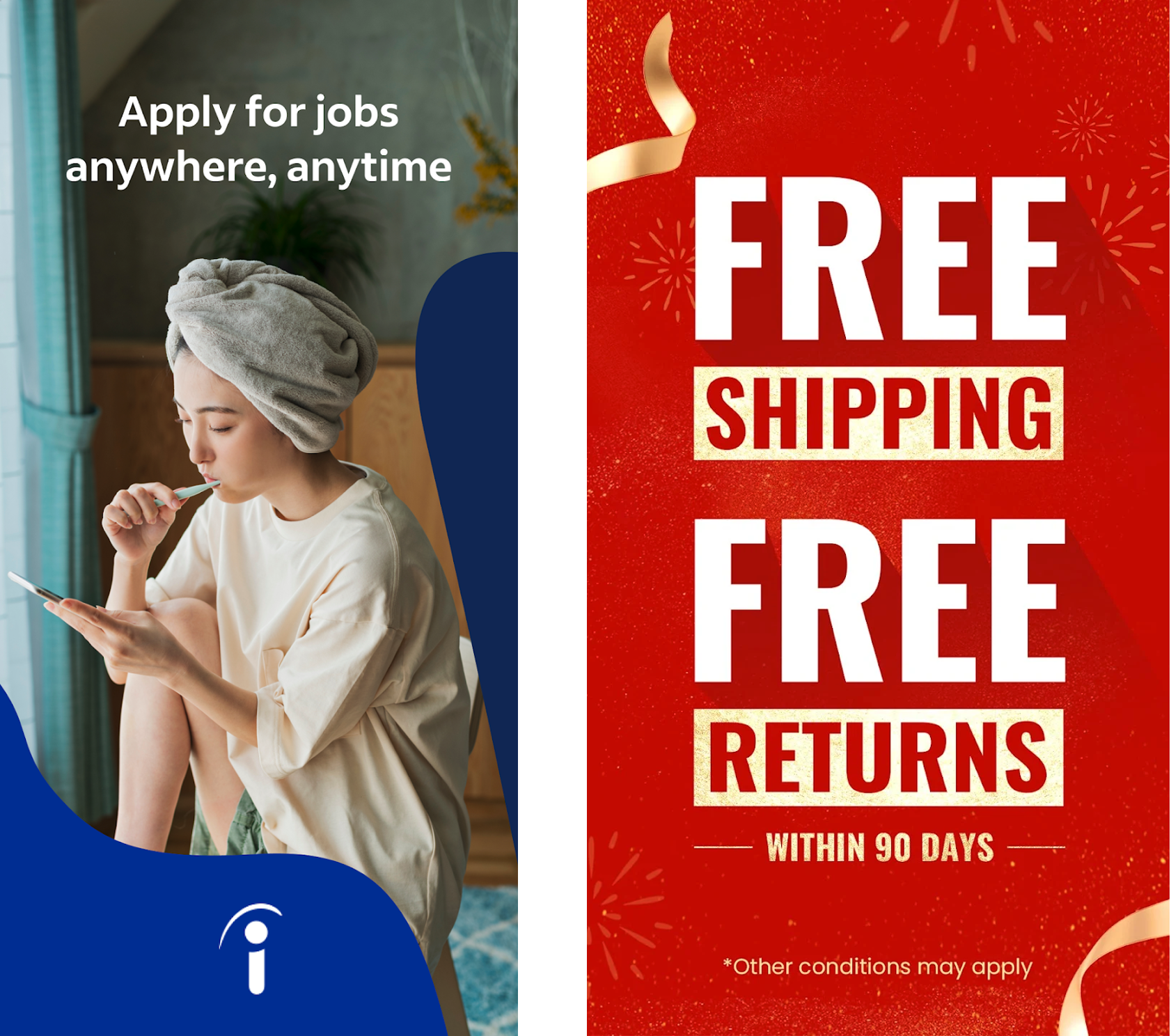}}
    
    \caption{Examples of app introduction images from Google Play app store.}
    \label{fig:app-introduction-images}
\end{figure*}

Google Play is one of the largest mobile app stores with thousands of apps from diverse domains. 
This makes it a rich source of inspiration for requirements elicitation and app design \cite{Ferrari:StrategiesBenefitsChallenges:2023, Maalej:AutomatedProcessingUser:2024}. 
Particularly, the app introduction images on Google Play are a gold mine for GUI retrieval, as they are carefully selected and described by developers to represent the important app features.
Figure \ref{fig:app-introduction-images} shows examples of these images.
A large portion can be categorised as \textit{surrounded screenshots}, each displaying a screenshot and a succinct caption that describes the screen (Figure \ref{fig:app-introduction-images} (a)).
In order to extract the screenshots and respective captions, we developed an automated \textbf{pipeline}, optimised for these images. 
The pipeline first classifies images into three categories as shown in Figure \ref{fig:app-introduction-images}. 
Then, from the \textit{surrounded screenshots} it crops the screenshot areas and extracts the captions (like ``LOG BLOOD PRESSURE'' in Figure \ref{fig:app-introduction-images} (a)).

By applying the pipeline on the introduction images of approximately 117k apps, we created a comprehensive dataset comprising 303k screenshots, of which 135k have captions. 
We refer to the collection of 303k screenshots as the \textbf{ScreenRepo} and the subset containing captions as \textbf{SCapRepo}.
We then used SCapRepo together with the Screen2Words and Clarity datasets to fine-tune the CLIP model and create a vision-language model specific to the GUI domain. 
We call the new model \textbf{GUIClip}. 
Our evaluation results demonstrate a promising performance of GUIClip for text-to-image GUI retrieval, outperforming both text-only approaches and the CLIP model. 
Based on GUIClip, we developed \textbf{GUing}, a search engine that accepts textual queries as input to retrieve relevant GUI images from our dataset as well as the Rico and ReDraw datasets. 
The engine can easily be extended with additional screenshots. 
Our manual evaluation shows that the search engine recommends highly relevant GUIs, which can serve as a valuable source of design inspiration and a tool for rapid prototyping and requirements refinement.
Two additional experiments suggest that GUIClip is also beneficial for other GUI-related tasks, such as sketch-to-GUI retrieval and GUI classification.
The paper provides the following contributions, with source code and datasets publicly available for research purposes \url{https://github.com/Jl-wei/guing}:

\begin{itemize}
\item A vision-language model named GUIClip for a range of GUI-related tasks, including text-to-GUI retrieval, sketch-to-GUI retrieval, and GUI classification available at \url{https://huggingface.co/Jl-wei/guiclip-vit-base-patch32}.

\item A GUI search engine that can achieve high performance with textual queries.

\item Two large GUI datasets containing 303k screenshots, 135k of which include captions.

\item An extensible pipeline for automatically extracting screenshots and captions from app introduction images.
\end{itemize}

\section{Dataset Creation}\label{sec:dataset-creation}
\begin{figure*}[!htb]
    \centerline{\includegraphics[width=1\textwidth]{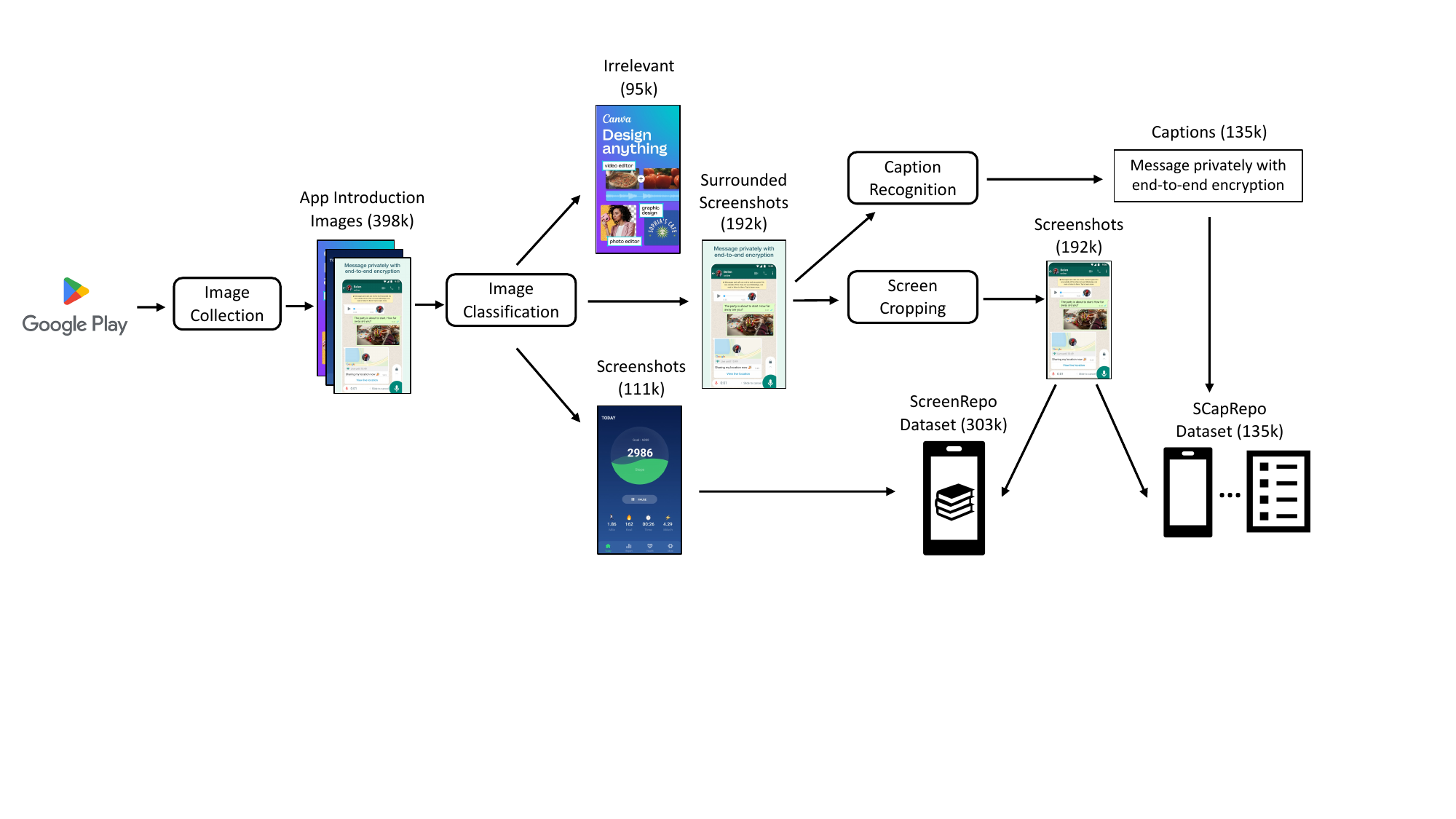}}
    \caption{Overview of dataset creation pipeline.}
    \Description{Overview of dataset creation pipeline.}
    \label{fig:dataset}
\end{figure*}

Created two datasets: (1) \textit{Google Play Screenshot Caption (SCapRepo)} including 135k screenshot-caption pairs for the training of our vision-language model and (2)
{Google Play Screenshot Repository (ScreenRepo)} including 303k screenshots for the search engine repository. 
Figure \ref{fig:dataset} depicts the datasets creation pipeline. 
In the initial step, we collected app introduction images from Google Play. 
Subsequently, we developed an image classifier to categorise the images into \textit{screenshots}, \textit{surrounded screenshots} and \textit{irrelevant}.
The area in a \textit{surrounded screenshot} containing a screenshots was precisely cropped from the surrounding by applying object detection.
Additionally, we extracted the captions from the \textit{surrounded screenshots} by applying Optical Character Recognition (OCR). 
The images classified as \textit{screenshots} together with the cropped screenshots constitute the ScreenRepo, while the screenshot-caption pairs extracted from \textit{surrounded screenshots} constitute the SCapRepo.
The following describes each step in detail.

\subsection{Image Collection}\label{sec:gui-image-collection}
Given the ID of an app, we can download the app introduction images via Google Play Scraper \cite{github-google-play-scraper}.
In order to obtain the app IDs, we initially gathered top ranked apps in each category from AppBrain \cite{AppBrain-app-ranking}, creating a seed list. 
An app on Google Play often suggests similar apps and the app developer who may also offer additional apps. 
This scenario can be conceptually treated as a graph, where apps represent nodes and the relationships (such as app similarity and common developer) represent edges.
Starting from the seed list, we performed a breadth-first search on this graph, collecting a total of 117,283 apps (by October 2023). 
All apps were collected from the US Google Play in English language.

The stylistic differences between the introduction images of games and non-game apps make the former unsuitable as reference for general app design. 
Consequently, we excluded games from our analysis, leaving a dataset of 85,631 non-game apps.
This includes a total of 874,130 images. 
Some of these images specifically demonstrates landscape views, Wear OS, or Android TV \cite{google-play-app-preview}.
To exclude those from our dataset, we applied a filtering process based on aspect ratio, leading to 553,755 precisely filtered images.
Additionally, some images could be duplicates despite having different filenames. 
To remove duplicates, we calculated the SHA-1 hash of each individual image, retaining only one image with the same hash. 
This yielded a final count of 398,511 unique app introduction images.

\subsection{Image Classification}
For subsequent analysis, we classified the images obtained from the prior phase into three categories as depicted in Figure \ref{fig:dataset}.
A subset of these images show the complete display of an app user interface: those are \textit{screenshots}. 
In some images, only a segment represents a captured screenshot. 
Those are categorised as \textit{surrounded screenshots}. 
The remaining images either do not show a screenshot or only a partial or a slanted screenshot. 
Those are classified as \textit{irrelevant}.
As a manual classification of around 400k images is unreasonable, we trained an image classifier to automatically classify these images into \textit{screenshots}, \textit{surrounded screenshot}, or \textit{irrelevant}.

\subsubsection{Classification Model}
We used the Vision Transformer (ViT) \cite{Dosovitskiy:ImageWorth16X16:2021} as classifier due to its top performance (as will be discussed in subsequent sections). 
The ViT architecture essentially constitutes the encoder of transformer as described by Vaswani et al.~\cite{Vaswani:AttentionAllYou:2017}. 
It has demonstrated superior performance in image classification tasks.
In the area of natural language processing, the dominant approach is to pre-train the model on a large text corpus followed by fine-tuning on a more targeted dataset \cite{Han:PretrainedModelsPresent:2021}. 
Similarly, the ViT benefits from a pre-training phase using a comprehensive collection of images, allowing the model to capture inherent image features. 
Consequently, when deployed on novel tasks, the ViT requires a reduced amount of task-specific training due to the knowledge acquired during its pre-training.

\subsubsection{Creation of Training Data}\label{sec:gui-classification-data}
To train and fine-tune the classification model, a labelled dataset is required.
For this, we initially sampled 5,000 random images from the collected filtered dataset (see Section \ref{sec:gui-image-collection}).
After eliminating duplicates, 3,615 unique images remained. 
To annotate these images, we created an annotation guide with examples to clarify the definition of \textit{screenshots}, \textit{surrounded screenshot}, and \textit{irrelevant}.
Two of the authors then annotated these images independently.
For the entire labelled dataset, only 7 images had conflicting labels, primarily due to disagreements between the \textit{surrounded screenshot} and \textit{irrelevant} categories. 
These conflicts arose because the images in question did not contain a full screenshot but rather a UI component, such as a widget.
After discussion, we reached a consensus to label images featuring a larger UI component (as thus potentially inspiring for a screen design) as \textit{surrounded screenshot}, and the others as \textit{irrelevant}.

\subsubsection{Classification Performance}\label{sec:classification-perf}
We evaluated the performance of the classifier using the labelled app introduction images from the previous step, which we split into a 80:20 ratio for training and testing, respectively.
The goal of training a machine learning (ML) model is to minimise the loss function, which quantifies the difference between the model's predicted output and the actual target values.
We trained the classification model using mini-batch gradient descent, with a batch size\footnote{Batch size refers to the number of data samples processed simultaneously during a single training step in ML.} of 64, AdamW \cite{Loshchilov:DecoupledWeightDecay:2019} optimiser\footnote{Optimisers, like AdamW, are strategies used to minimise the loss function.}, and an initial learning rate\footnote{Learning rate is a parameter in ML algorithms, controlling the training step size while moving toward a minimum of a loss function.} of $2e^{-5}$.
The classifier was trained in 5 epochs on a machine with a NVIDIA Tesla T4 GPU with 16 GB VRAM.
To enhance the robustness of our results, we performed a 10-fold cross-validation by repetitively splitting the dataset into distinct training and testing subsets ten times in a randomised manner. 
We then computed the mean values of the \textit{precision, recall}, and \textit{F1 score} across all runs.
The results presented in Table \ref{tab:classification-acc} show a top performance of our classifier, with an average F1 score of 0.964.

\begin{table}[]
\centering
\caption{Accuracy of the image classification (mean values for 10-fold cross-validation).}
\begin{tabular}{l | ccc}
\toprule
Class & Precision & Recall & F1 \\
\midrule
Surrounded Screenshot & 0.974 & 0.978 & 0.976 \\
Screenshot & 0.967 & 0.970 & 0.968 \\
Irrelevant & 0.937 & 0.926 & 0.931 \\
\textbf{Weighted Average} & \textbf{0.964} & \textbf{0.964} & \textbf{0.964}\\
\bottomrule
\end{tabular}
\label{tab:classification-acc}
\end{table}

\subsubsection{Applying the Classifier}\label{sec:gui-classification-appl}
Subsequent to the evaluation, we trained a ViT model on all of the 3,615 images. 
This model was then used to classify all of the filtered app introduction images in the dataset.
To increase the reliability and integrity of our repository, we calibrated the classifier's threshold at a high confidence level of 0.9, which means that only images with a classification probability exceeding  0.9 are included in the respective category.
Our empirical investigation reveals that, at the 0.9 threshold, the precision scores for the categories \textit{screenshots} and \textit{surrounded screenshots} surpass 0.99. 
Finally, our repository had 111,847 images classified as \textit{screenshots} and 191,993 as \textit{surrounded screenshots}.

\subsection{Screen Cropping}
The \textit{surrounded screenshots} are the images that contain screenshots with additional visual frames.
To automatically crop the screenshot areas from these \textit{surrounded screenshots}, we trained an object detection model. 
This model is capable of localising the screenshot within the larger image and subsequently generating a bounding box. 
A bounding box is defined as a rectangular delineation that encases the area of interest: in this case, the screenshot. 
Given the bounding boxes inside the \textit{surrounded screenshots}, we can easily crop the screenshot area using an image processing library.

\subsubsection{Object Detection Model}
We use DETR (DEtection TRansformer) \cite{Carion:EndtoEndObjectDetection:2020} as screenshot detector.
DETR is an end-to-end object detector, composed of a Convolutional Neural Network (CNN) backbone and a encoder-decoder transformer \cite{Vaswani:AttentionAllYou:2017}.
Atop the transformer decoder, dual output heads are appended: a linear layer dedicated to categorise class labels and a multi-layer perceptron (MLP) tasked with the generation of bounding boxes for the object location.
The model uses so-called object queries to detect objects in an image. 
Each object query is designed to search specifically for one particular object of interest in the given image.
We set the number of queries to 1, as we are interested in the largest screenshot within a \textit{surrounded screenshot}.

\subsubsection{Creation of Training Data}
We created a separate dataset to fine-tune the detection model for screenshot cropping. 
From the dataset created in Section \ref{sec:gui-classification-data}, we collected all the \textit{surrounded screenshots} and labelled the screenshot areas with a bounding box.
The annotation was performed with Prodigy \cite{prodigy}.
For each image, an author drew one bounding boxes that cover the screenshot. 
The results were reviewed by another author.
The majority of the \textit{surrounded screenshots} feature only a single screenshot.
In instances where a single image contains multiple screenshots, annotation was selectively applied only to the most prominent (largest) screenshot. 
This resulted in a dataset comprising 1,768 annotated images, each including a delineated bounding box that specifies the screenshot area.

\subsubsection{Detection Performance}
Intersection over Union (IoU) is a standard metric for object detection tasks. 
It evaluates the extent of overlap between the predicted bounding box and the ground truth bounding box.
If $IoU = 0$, it means that there is no overlap between the boxes, while $IoU = 1$ means that the overlap is perfect.
\[ \mathit{IoU} = \frac{Area Of Overlap}{Area Of Union} \]
We followed the evaluation protocol of the Common Objects in Context (COCO) \cite{Lin:MicrosoftCOCOCommon:2014}, which is a common object detection benchmark.
As stated in the protocol, we quantitatively evaluated the object detection performance by calculating precision and recall metrics at various IoU thresholds. 
Specifically, we computed the precision values at three IoU levels: 0.50, 0.75, and an average over the range from 0.50 to 0.95. 
Similarly, recall was determined over the IoU range of 0.50 to 0.95.

We used 80\% of the data for training and the remaining 20\% for evaluation.
The detector was trained on the training set using mini-batch gradient descent, with a batch size of 16 and AdamW optimiser with an initial learning rate of $1e^{-5}$ and ten training epochs.
Table \ref{tab:iou-acc} shows the results for a 10-fold cross-validation, indicating a top performance.

\begin{table}[]
\centering
\caption{Accuracy of the screen localisation (mean values for 10-fold cross-validation).}
\begin{tabular}{l | ll}
\toprule
\multirow{3}{*}{Precision} & IoU=0.50:0.95 & 0.919 \\
                  & IoU=0.50 & 0.981 \\
                  & IoU=0.75 & 0.971 \\
                  \midrule
Recall            & IoU=0.50:0.95 & 0.947 \\
\bottomrule
\end{tabular}
\label{tab:iou-acc}
\end{table}

\subsubsection{Applying the Detector}\label{sec:gui-detection-appl}
We trained a DETR model with all of the 1,768 labelled images, and applied this model to the 191,993 images categorised as \textit{surrounded screenshot}.
DETR predicted a bounding box that localise the screenshot for each mage. 
We then cropped these images according to the bounding box.
The screenshots cropped from the \textit{surrounded screenshot} as well as the app introduction images previously classified as \textit{screenshots} represent the ScreenRepo dataset.

\subsection{Caption Recognition}
The majority of the images, classified as \textit{surrounded screenshots}, include captions that provide a succinct descriptions of the screenshots. 
Typically, these captions are positioned either above or below the image. 
In order to accurately extract this textual information, we employed Optical Character Recognition (OCR), a commonly used technology for recognising text displayed within images.

\subsubsection{Applying OCR}\label{sec:gui-ocr-appl}
We used PaddleOCR \cite{Li:PPOCRv3MoreAttempts:2022}, which is an industrially robust OCR model renowned for its accuracy and swift performance.
The application of PaddleOCR is adequate since the textual content within the images under analysis is markedly legible and usually not hand-written.
PaddleOCR processes an input image and outputs a collection of identified elements.
Each element is comprised of a bounding box delineating the text area, its recognised text, and an associated confidence level reflecting the recognition certainty. 
We applied PaddleOCR on all of the 191,993 \textit{surrounded screenshots}. 
For each \textit{surrounded screenshot}, text bounding boxes that exhibit spatial overlap with the screenshot's bounding box, as delineated in Section \ref{sec:gui-detection-appl}, were excluded. 
The remaining text bounding boxes were subsequently aggregated to form a coherent caption for the corresponding image.

\subsubsection{Post-processing}
Some of the \textit{surrounded screenshots} may not contain captions. 
These images were removed from our analysis.
The captions extracted are not exclusively in English. 
They encompass a variety of languages including French, German, and Arabic, among others.
As we focus on English in this work, captions in languages other than English were excluded.
To identify the language of each caption, we employed Lingua \cite{github-lingua}, an effective language detection tool.
Furthermore, to correct any spelling error within the captions due to potential OCR errors, we utilised Autocorrect \cite{github-autocorrect}, a Python-based spelling corrector.

It is noteworthy that in some instances, multiple \textit{surrounded screenshots} of an app may feature identical captions.
This suggests that the caption is either overly generic or simply a logo of the app, and thus may not adequately convey the screenshot content. 
To address this redundancy, we conducted a filtering process whereby, for each app, only the first \textit{surrounded screenshot} from the set containing identical captions was retained.
We observed that the first \textit{surrounded screenshot} displayed by an app on Google Play typically showcases the most representative feature.
The post-processing finally left 135,357 screenshot-caption pairs, collectively referred to as the SCapRepo dataset.

\section{GUI Search Engine}
\begin{figure*}
    \centering
    \centerline{\includegraphics[width=0.9\textwidth]{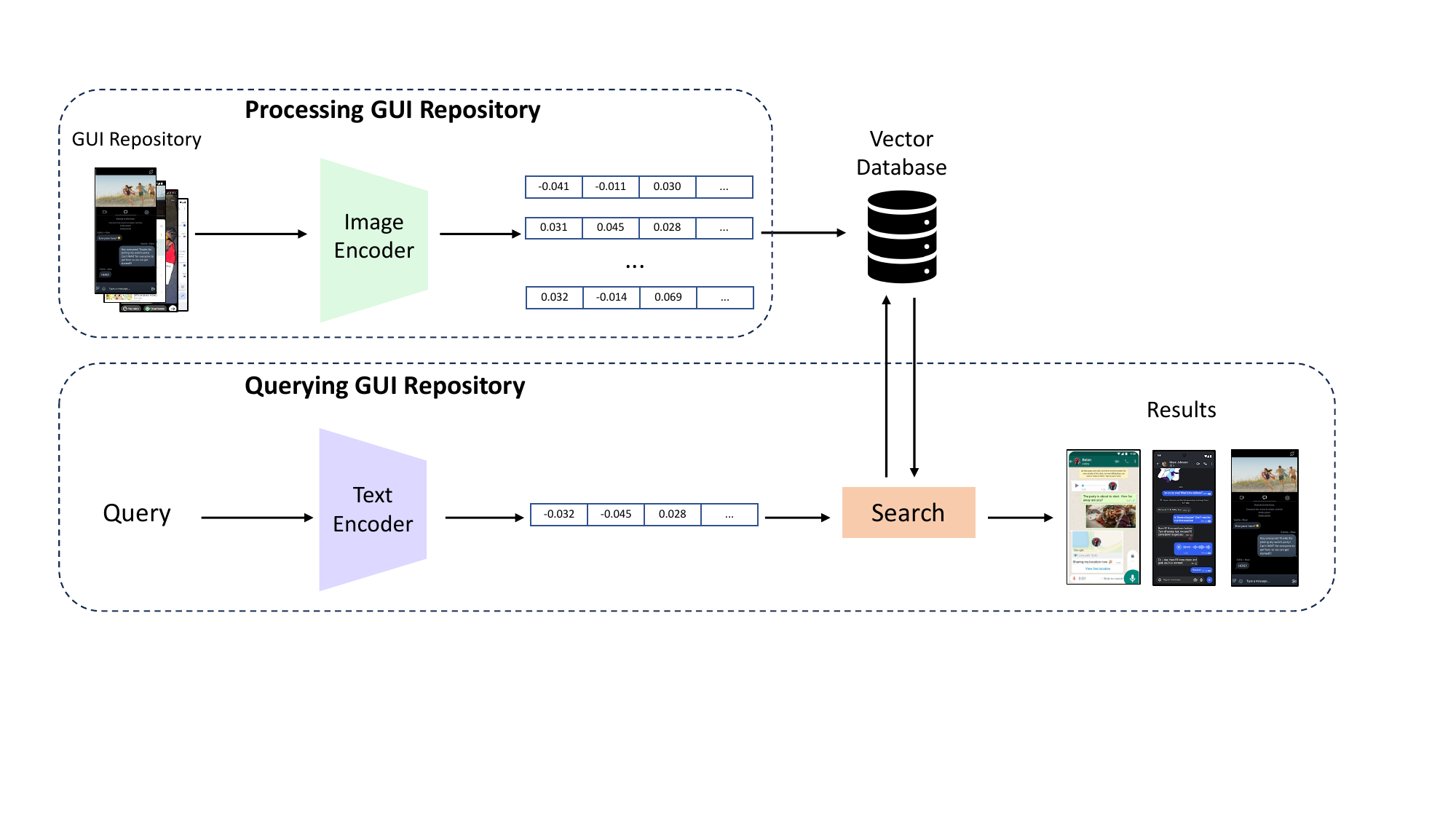}}
    \caption{Overview of GUing, our GUI search engine for text-to-GUI retrieval.}
    \Description{Overview of GUing, our GUI search engine for text-to-GUI retrieval.}
    \label{fig:search-engine}
\end{figure*}

We introduce GUing, an advanced search engine based on a vision-language model that leverages textual queries to retrieve relevant screenshots from a large GUI repository which consists of ScreenRepo, Rico \cite{Deka:RicoMobileApp:2017} and ReDraw \cite{Moran:MachineLearningBasedPrototyping:2020}.
The architecture of GUing is illustrated in Figure \ref{fig:search-engine}.
GUing is based on the GUIClip model.
The engine uses the image encoder and text encoder modules of GUIClip to embed screenshots and textual queries within a unified latent space.
This enables an efficient search for the screenshots by calculating the cosine similarity between the text query embedding and the screenshot embedding.
In the following, we introduce the details of GUIClip and GUing.

\subsection{Constructing the GUIClip Model}
GUIClip, our vision-language model, serves as the fundamental component of the GUI search engine, integrating the image and text modalities. 
Building on CLIP \cite{Radford:LearningTransferableVisual:2021}, GUIClip offers enhanced capabilities in multimodal representation learning for the GUI domain.
CLIP (Contrastive Language–Image Pre-training) is a significant foundation model in multimodal representation learning trained on a large-scale image-caption dataset. 
In this work, we construct a GUI-specific CLIP model by training a vision-language model on a large-scale dataset of screenshot-caption pairs. 

\subsubsection{Training Data and Pre-processing}\label{sec:GUIClip-data}
To train GUIClip learn the visual representation of mobile screenshots, we combined three datasets:

\begin{itemize}
\item Screen2Words dataset \cite{Wang:Screen2WordsAutomaticMobile:2021} is a subset of Rico. 
It contains 112,085 textual summaries for 22,417 unique screenshots.
The screenshot summaries are written by professional annotators.
For each screenshot, five summaries are created by five different annotators.

\item Clarity dataset \cite{Moran:EmpiricalInvestigationUse:2022} consists of 45,998 descriptions for 10,204 screenshots from popular Android apps.
Crowd workers summarised each screenshot in different granularity, with one high-level caption and up to four low-level detailed descriptions.

\item Google Play Screenshot Caption (SCapRepo) dataset, where textual descriptions are created by developers and app vendors.  
The dataset includes 135k screenshot-caption pairs (see Section \ref{sec:dataset-creation}).
The screenshots are cropped from app introduction images on Google Play. 
The corresponding captions briefly describe the screenshots.
\end{itemize}

The captions were first tokenised with the CLIP tokeniser \cite{github-clip}.
Captions longer than 77 tokens were truncated to 77, captions shorter were padded to 77.
To process the screenshot images, we resized the resolution to $224 \times 224$. 
Furthermore, we proportionately rescaled the pixel value of the images from a range of 0-255 to 0-1. 
This new value was then normalised using the parameters (means and standard deviations) featured in the official CLIP documentation \cite{github-clip}.

\subsubsection{Model Architecture}
GUIClip consists of two components, an image encoder and a text encoder.
The image encoder, which adopts a Vision Transformer (ViT) model \cite{Dosovitskiy:ImageWorth16X16:2021}, ingests image data and generates its corresponding image embedding. 
The text encoder is a transformer encoder \cite{Vaswani:AttentionAllYou:2017} which accepts text as input to produce an associated text embedding.
Employing contrastive learning techniques, the image and text embeddings are mapped into a multimodal embedding space. 
In this shared space, images and texts presenting semantic similarity are mapped close to each other.

Contrastive learning on image-text pairs constructs a bridge between visual perception and natural language. 
Given a batch of N (image, text) pairs, the training objective is to identify the authentic pairings from $N \times N$ potential (image, text) combinations within the batch \cite{Radford:LearningTransferableVisual:2021}.
To achieve this objective, the model formulates a multimodal embedding space, concurrently training an image encoder and text encoder to maximise the cosine similarity of the image and text embeddings derived from the N authentic pairs in the batch. 
Simultaneously, it minimises the cosine similarity of the embeddings from the remaining $N^2-N$ incorrect pairings.

\begin{figure}
    \centering
    \centerline{\includegraphics[width=0.7\textwidth]{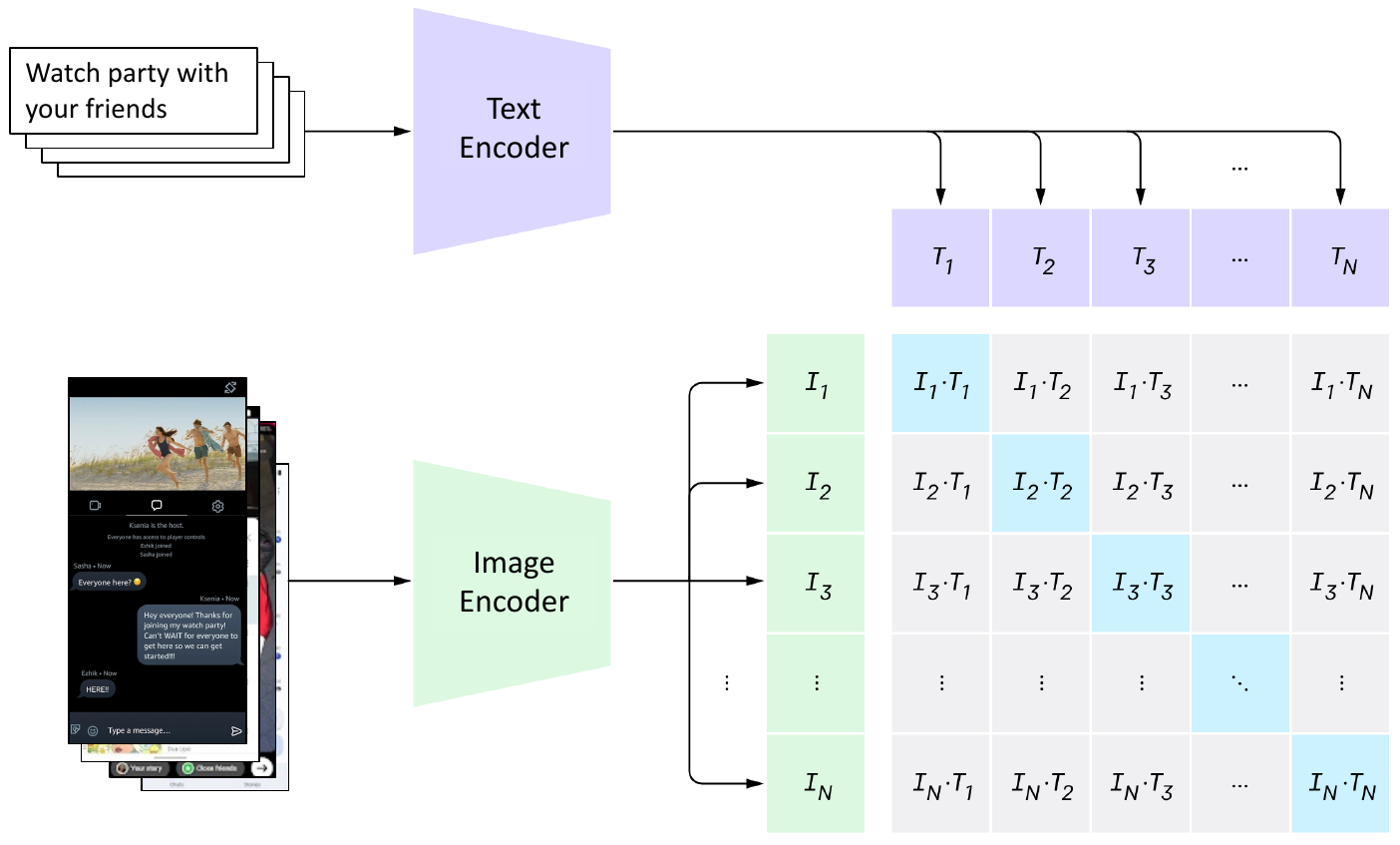}}
    \caption{Overview of contrastive learning to train our vision-language model GUIClip (adapted from Radford et al.~\cite{Radford:LearningTransferableVisual:2021}).}
    \Description{Overview of contrastive learning to train our vision-language model GUIClip (adapted from Radford et al.~\cite{Radford:LearningTransferableVisual:2021}).}
    \label{fig:GUIClip}
\end{figure}

\subsubsection{Training Details}
In order to adapt the CLIP model to GUIClip, rather than training the CLIP model from scratch where the weights of image encoder and text encoder are randomly initialised, our training is based on the checkpoint of "openai/clip-vit-base-patch32" \cite{huggingface-clip}.
This checkpoint has been pre-trained on a 400 million image-text pairs.
This method allows us to harness the benefits inherent to pre-trained models and, as a result, substantially diminishes the time required for training. 
The model was trained on all of the collected screenshot-caption pairs for 5 epochs.
We applied mini-batch gradient descent with a batch size of 128 and AdamW optimiser with an initial learning rate of $5e^{-5}$.

\subsection{Searching the GUI Repository}

\subsubsection{Processing the GUI Repository}
The GUI repository is composed of 383k screenshots from ScreenRepo (303k), Rico (66k) and ReDraw (14k).
These screenshots are stored in image formats such as .jpeg and cannot be directly queried.
Therefore, we embedded all the screenshots in the repository and saved them in a vector database.

The screenshots are first processed (i.e.~resized, rescaled, and normalised) with the same parameters discussed in Section \ref{sec:GUIClip-data}.
The processed screenshots subsequently serve as input for the image encoder of GUIClip.
The image encoder accepts an image as an input and returns a 512-dimensional embedding. 
The entire GUI repository, comprising 383k screenshots, is processed via the image encoder, consequently generating 383k 512-dimensional embeddings.
The mapping between the screenshot ids and their embeddings are saved in a vector database for the retrieval.

\subsubsection{Querying the GUI Repository}
In the querying phase, the textual query is encoded by GUIClip's text encoder, resulting in a 512-dimensional embedding. 
Subsequently, this query embedding is utilised to identify the top-k most similar screenshot embeddings from the vector database.
We use cosine similarity as the measure of similarity. 
Consequently, the query results consist of the top-k most similar screenshots.

\subsubsection{Accelerating the Searching}
Given the large scale of the GUI repository, employing a brute force approach to compare the query embedding with all screenshot embeddings is inefficient.
To tackle this problem, we applied approximate nearest neighbour search \cite{Arya:OptimalAlgorithmApproximate:1998} to accelerate the querying. 
To this end, we established \textit{n} Voronoi cells within the embedding space, with each screenshot embedding falling into one of these designated cells. 
During a search, the query embedding is initially compared with the centroid embeddings of each cell.
This method results in a considerable reduction of necessary comparisons when contrasted with the total number of screenshot embeddings.
The \textit{k} cells with centroids closest to the query embedding are identified, whereupon the screenshot embeddings within these cells are thereafter compared with the query embedding in order to pinpoint similar screenshot embeddings.
Our implementation is based on Faiss \cite{douze2024faiss}, a library for efficient similarity search.
We set the number of cells \textit{n} to 3000, and the number of cells that are visited to perform a search \textit{k} to 1000.

\section{Evaluation Design}
For the empirical evaluation of our search engine as well as our vision-language model, we focus on answering the following research questions:

\begin{itemize}
\item \textbf{RQ1: How does GUIClip perform in text-to-GUI retrieval tasks compared to state-of-the-art approaches?}
As the CLIP model \cite{Radford:LearningTransferableVisual:2021} was trained on 400 millions of image-caption pairs from the internet, it can also be applied on text-to-GUI retrieval task.
An alternative method involves leveraging only the textual information: i.e.~the text displayed on the GUI images, computing its semantic similarity with the input textual query.
Thus, the question is whether the performance our proposed GUIClip model surpasses the CLIP model and text-only approaches.

\item \textbf{RQ2: How relevant are the retrieved screenshots for queries representing app features?}
As a search engine, GUing processes textual queries as input and searches for relevant images within the GUI repository. 
It is thus important to evaluate how effective is GUing in delivering useful results that actually align with user expectations.

\item \textbf{RQ3: Does GUIClip also show a high performance in other GUI-related tasks?}
The CLIP model, as a foundation model, has been extensively applied across various vision-language tasks, consistently delivering promising results \cite{Zhang:VisionLanguageModelsVision:2024}. 
As GUIClip is a fine-tuned version specifically with GUI images, it has the potential to perform well for other GUI-related tasks. 
Particularly, the question is whether GUIClip can also outperform CLIP in other GUI-related tasks.
\end{itemize}

To answer these research questions, we designed four experiments. 
The first two experiments evaluate our approach on text-to-GUI retrieval:  the first experiment (Exp1) addresses RQ1 by benchmarking three screenshot-caption datasets, while the second (Exp2) addresses RQ2 through a manual evaluation of search engines. 
The final two experiments focus on RQ3, assessing the performance of GUIClip on more GUI-related tasks:
the third experiment (Exp3) focusing on GUI classification and the fourth (Exp4) on sketch-to-GUI retrieval.

\subsection{Exp1: Evaluation of GUIClip for Text-to-GUI Retrieval}
In the first experiment, the text-to-GUI retrieval performance of our GUIClip model is evaluated together with baseline models on three distinct datasets.
During the evaluation, every screenshot and its associated captions in test dataset are embedded using the models under evaluation.
For every single caption, all screenshots within the test dataset are ranked by the cosine similarity of the caption embedding and screenshot embedding.
A comparative analysis of the performance of GUIClip against various baselines provide answers to RQ1.

\subsubsection{Experimental Data}
From each of the screenshot datasets described in Section \ref{sec:GUIClip-data} (SCapRepo, Screen2Words, and Clarity), we randomly selected 1000 screenshots for validation, 1000 screenshots for test, and the rest for model training, as shown in Table \ref{tab:gui-dataset-split}.
Similar to the approach suggested by Wang et al.~\cite{Wang:Screen2WordsAutomaticMobile:2021}, we split the data to \textit{not} share the screenshots from the same app across different splits. 
That is, all the apps and screenshots in the test and validation set were completely unseen during the training. 
This arrangement allows us to assess how well our model generalises to previously unseen screenshots from unseen apps during the test phase.

\begin{table}[]
\centering
\caption{Statistics of the evaluation datasets.}
\begin{tabular}{l | cccc}
\toprule
Split & Dataset & \#Apps & \#Screenshots & \#Captions \\
\midrule
\multirow{3}{*}{Training} & SCapRepo & 32720 & 133477 & 133477 \\
& Screen2Words & 5684 & 20379 & 101895 \\
& Clarity & 3346 & 8204 & 36964 \\
\midrule
\multirow{3}{*}{Validation} & SCapRepo & 246 & 1000 & 1000 \\
& Screen2Words & 281 & 1000 & 5000 \\
& Clarity & 413 & 1000 & 4475 \\
\midrule
\multirow{3}{*}{Test} & SCapRepo & 245 & 1000 & 1000 \\
& Screen2Words & 286 & 1000 & 5000 \\
& Clarity & 416 & 1000 & 4559 \\
\bottomrule
\end{tabular}
\label{tab:gui-dataset-split}
\end{table}

\subsubsection{GUIClip Training Details}
We trained the "openai/clip-vit-base-patch32" \cite{huggingface-clip} checkpoint on the training set for a total of 5 epochs. 
During the training, a mini-batch gradient descent approach was employed with a batch size of 128. 
To optimise the learning process, the AdamW optimiser was utilised with an initial learning rate set to $5e^{-5}$.

\subsubsection{Baselines}\label{sec:rq1-baseline}
As highlighted by Bernal-Cardenas et al.~\cite{Bernal-Cardenas:GuigleGUISearch:2019} and by Kolthoff et al.~\cite{Kolthoff:DatadrivenPrototypingNaturallanguagebased:2023}, existing approaches perform their queries based on GUI metadata, which is absent in screenshots collected by processing app introduction images from Google Play. 
Consequently, we implemented a text-only retrieval approach by ourselves.
In addition, we compared our model with (a) the original CLIP model without further fine-tuning and (b) with the CLIP model fine-tuned \textit{without} our SCapRepo data.
This results in three baselines: 

\begin{itemize}
\item \textit{OCR + BGE (Text Only)}:
As discussed in the Section \ref{sec:gui-ocr-appl}, PaddleOCR \cite{Li:PPOCRv3MoreAttempts:2022} is an industrial grade OCR library.
BGE \cite{Xiao:CPackPackagedResources:2023} is a state-of-the-art text embedding model, trained in three steps: pre-training with plain texts, contrastive learning on text pair dataset, and task-specific fine-tuning.
First, the screenshots are embedded by using OCR + BGE in three steps: 1) texts displayed on the GUI images are extracted with PaddleOCR; 2) the extracted text is then concatenated into a sentence, separated by semicolons; 3) the sentence is then embedded by BGE.
Finally, the captions are directly embedded using the BGE model.

\item \textit{CLIP}:
The screenshots are embedded by the image encoder of CLIP and the captions are embedded by the text encoder of CLIP.
The "openai/clip-vit-base-patch32" checkpoint is used for this evaluation.

\item \textit{GUIClip-CS}:
This is a fine-tuned model from CLIP using the Clarity and Screen2Words dataset.
The training process is exactly the same as for GUIClip, except that our SCapRepo dataset is excluded during the training.
The image encoder of GUIClip-CS embeds the screenshots, while its text encoder embeds the captions.
\end{itemize}

\subsubsection{Evaluation Metric}\label{sec:eval-metric-recall}
The recall@k is commonly used for evaluating text-to-image retrieval performance \cite{Radford:LearningTransferableVisual:2021,Singh:FLAVAFoundationalLanguage:2022,Li:BLIPBootstrappingLanguageImage:2022,Li:BLIP2BootstrappingLanguageImage:2023}.
Recall@k is defined as the percentage of captions whose corresponding screenshots fall into its top-k most similar screenshots.
For our evaluation, we used recall@1, recall@3, recall@5, recall@10, recall@50 and recall@100.

\subsection{Exp2: Manual Evaluation of GUing}
\label{sec:exp2}
A GUI search engine accepts textual queries as input and identifies the corresponding images within the GUI repository as output.
Exp1 focuses on evaluating the performance of retrieval models on the test datasets. 
Thereby, the main limitation is the use of the screenshot-caption pairs as ground truth, assuming a direct one-to-one correspondence between a single caption (search query) and a relevant screenshot.
However, such assumption contradicts real-world scenarios where one query can pertain to multiple screenshots.

To address this limitation, our second experiment aims to evaluate different search engines composed of various retrieval models and GUI repositories, through a manual assessment.
Specifically, we measure the proportion and ranks of images retrieved by the GUI search engines that are relevant to the query entered by the user.
Through a comparative analysis of the results usefulness between GUing and baseline approaches, we are then able to address RQ2.

\subsubsection{Baselines} 
To answer RQ2, we compared our search engine to two baselines: RaWi and GUing-CS.

\begin{itemize}
\item \textit{RaWi}.
RaWi \cite{Kolthoff:DatadrivenPrototypingNaturallanguagebased:2023} is a data-driven GUI prototyping approach that retrieves screenshots for reuse from Rico dataset based on natural language searches.
The search engine of RaWi is a BERT-based \cite{Devlin:BERTPretrainingDeep:2019} Learning-To-Rank (LTR) model that is trained on GUI relevance data collected from crowd-workers.
As Rico dataset provides the metadata of the screenshots, RaWi creates the textual representation of the screenshots by assimilating elements such as the activity names, UI component text, resource ids, and icon ids.
This integrated textual representation is then compared with the user text query, forming the basis for the GUI ranking.

\item \textit{GUing-CS}.
The architecture of GUing-CS replicates that of GUing. 
However, the retrieval model of GUing-CS is GUIClip-CS, which only uses the Clarity and Screen2Words datasets for training. 
That is, we explicitly remove our SCapRepo dataset from the training process. 
Furthermore, the GUI repository for GUing-CS is comprised exclusively of the Rico and ReDraw datasets, without the ScreenRepo dataset.
\end{itemize}

\subsubsection{Evaluation Method}
\begin{figure}
    \centering
    \centerline{\includegraphics[width=0.7\textwidth]{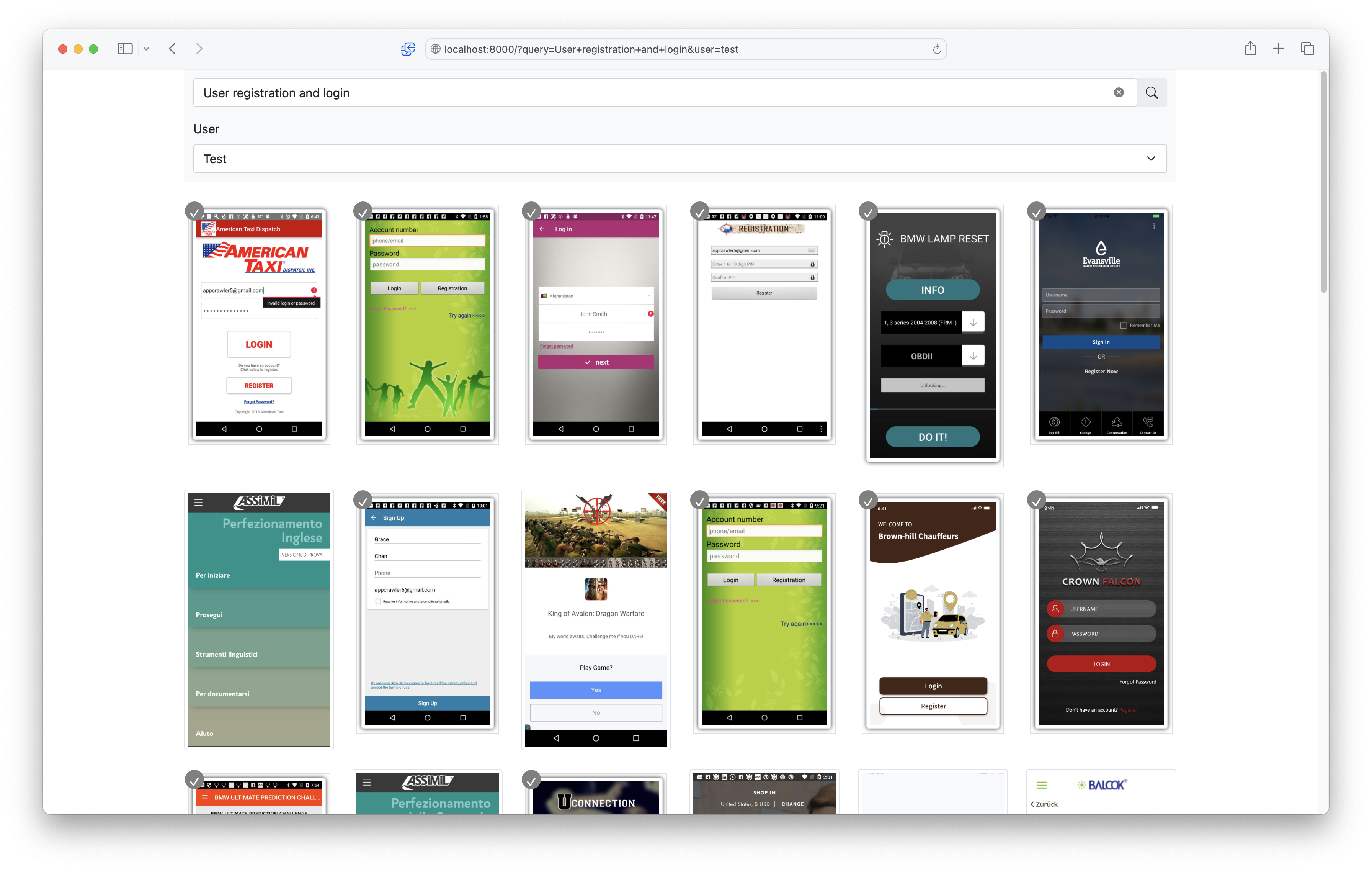}}
    \caption{Interface of the evaluation tool. It displays repeated GUIs due to their retrieval by more than one search engine.}
    \Description{Interface of the evaluation tool. It displays repeated GUIs due to their retrieval by more than one search engine.}
    \label{fig:eval-tool}
\end{figure}

To compare the search engines, we developed a tool as shown on Figure \ref{fig:eval-tool}.
The tool accepts textual queries as input and produces the top-10 screenshots from the three search engines: GUing, GUing-CS, and RaWi.
The tool mixes and randomly shuffles the resulting 30 screenshots to prevent possible biases by the evaluators. 
The origin of each search results is unclear and cannot be guessed by the evaluator. 
As shown on Figure \ref{fig:eval-tool}, GUIs might be displayed multiple times in the result due to their retrieval by more than one search engine.

During the assessment, the evaluators are asked to search for provided text queries, select all screenshots they consider relevant and confirm their selections via the submission button.
A screenshot is relevant to a query if it either displays a UI implementation of the feature delineated in the query or if it can serve as a reference or inspiration for the development of that feature.

The evaluation queries were derived from articles by AttractGroup and BuildFire, which are two companies specialised in mobile app development services and have substantial knowledge in this field.
AttractGroup's article enumerates 138 mobile app features to be considered during mobile app development, consisting of 45 general features and 93 domain specific features spanning across 11 domains (such as E-learning, mHealth, FinTech, etc.) \cite{138AppFeatures}.
BuildFire's article proposes 50 innovative app ideas for 2024, forming a good source of inspiration for prospective entrepreneurs \cite{50AppFeatures}.
From these two resources, we collected a total of 188 mobile app features. 
These were then divided into two sets: one set consisting of the 45 general and 50 innovative features, and another set comprising 93 domain-specific features.

Four evaluators took part in this experiment, each holding a master  degree in computer science and having a minimum of five years of software development experience.
Two of them performed the evaluation with the first set of queries, the other two used the second set of queries. 
This ensured that each query was evaluated independently by two evaluators.
The evaluators performed the search and submitted for their respective results as described above.

\subsubsection{Evaluation Metric}
To measure the performance of the GUI search engine, we applied the common information retrieval metrics Precision@k, Mean Reciprocal Rank, and HITS@k (similar to Kolthoff et al.~\cite{Kolthoff:DatadrivenPrototypingNaturallanguagebased:2023}):
\begin{itemize}
\item \textit{Precision at k (P@k)} quantifies the proportion of retrieved screenshots considered relevant, denoted as $|R_k|$, with respect to the top-k retrieved GUIs, denoted as $k$.
\begin{equation}
P@k = \frac{|R_k|}{k}
\end{equation}

\item \textit{Mean Reciprocal Rank (MRR)} is a measure for evaluating any process that produces an ordered  list of possible responses to a sample of queries. 
$rank_i$ signifies the rank of the highest ranked relevant screenshot of the i-th query. 
$Q$ stands for the total number of queries.
\begin{equation}
MRR = \frac{1}{Q}\sum_{i=1}^{|Q|}\frac{1}{rank_i}
\end{equation}

\item \textit{HITS@k} is a binary measure that validates the presence of at least one relevant GUI among the top-k ranked screenshots. 
It has the value 1 if there is at least one relevant screenshot within the top-k, otherwise, it is 0.
\begin{equation}
HITS@k = \begin{cases}
\; 1 \; |R_k| > 1\\
\; 0 \; |R_k| = 0
\end{cases}
\end{equation}
\end{itemize}

\subsection{Exp3: Evaluation of GUIClip for GUI Classification}
GUI classification aims to categorise a screenshot as a whole under a specific label.
It can be a fundamental activity for other GUI-related tasks, like GUI captioning \cite{Leiva:DescribingUIScreenshots:2023}, the tagging of user screenshots (e.g.~submitted in support requests) with certain requirements, components, or devices \cite{Maalej:WhenUsersBecome:2009, Roehm:MonitoringUserInteractions:2013, Martens:ExtractingAnalyzingContext:2019}, as well as an early prototyping by a larger group and multiple screens \cite{Pham:FirstImplementationDesign:2018}.
In this section, we explore the performance of GUIClip on GUI classification under zero-shot and linear-probe setting using the Enrico dataset \cite{Leiva:EnricoDatasetTopic:2020}.

\begin{figure}
    \centering
    \centerline{\includegraphics[width=1\textwidth]{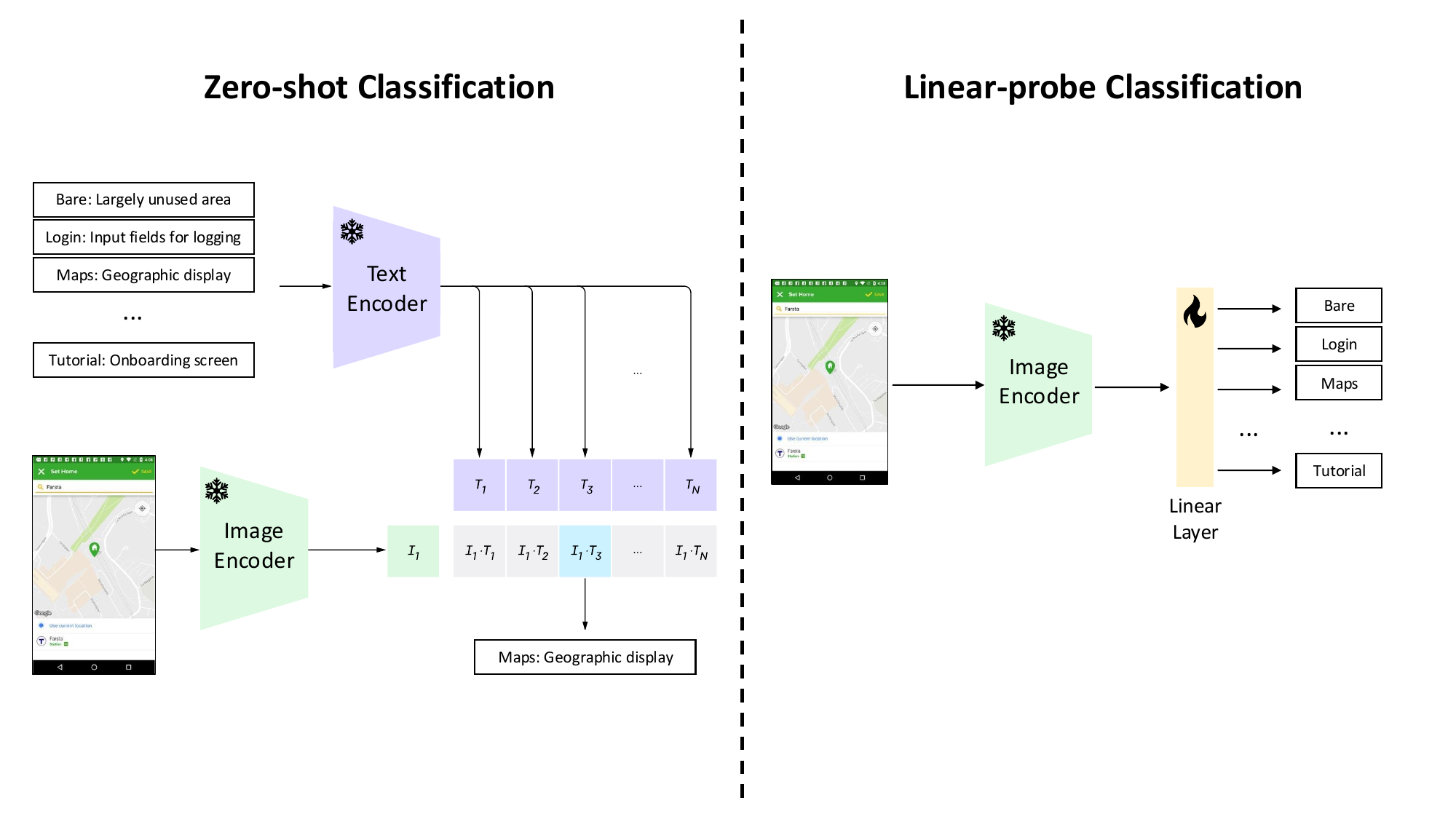}}
    \caption{Overview of zero-shot and linear-probe classification on Enrico dataset with GUIClip (left side adapted from Radfort et al.~\cite{Radford:LearningTransferableVisual:2021}).}
    \Description{Overview of zero-shot and linear-probe classification on Enrico dataset with GUIClip (left side adapted from Radford et al.~\cite{Radford:LearningTransferableVisual:2021}).}
    \label{fig:enrico-classification}
\end{figure}

\subsubsection{Model Architecture}
Like Radford et al.~\cite{Radford:LearningTransferableVisual:2021}, we evaluated the performance of GUIClip for GUI classification under zero-shot and linear-probe settings as shown on Figure \ref{fig:enrico-classification}.
Under both settings, the image encoder is frozen. 
\begin{itemize}

\item \textit{Zero-shot}.
In the zero-shot setting, all classification labels are embedded to text embedding with the text encoder.
During the classification process, the screenshot image is encoded by the image encoder, and the resulting screenshot embedding is compared to all text label embeddings.
The label exhibiting the highest similarity to the GUI image embedding is subsequently declared as the predicted outcome.

\item \textit{Linear-probe}.
In the linear-probe setting, we added a linear layer of shape $m \times k$ to the output of GUIClip image encoder.
\textit{m} refers to the output dimension of image encoder, while \textit{k} is the number of classification labels.
\end{itemize}

\subsubsection{Baselines}
We compared GUIClip performance with two baselines for zero-shot and linear-probe classification:

\begin{itemize}

\item \textit{CLIP}.
We also evaluated the performance of CLIP on GUI classification.
The setting of zero-shot and linear-probe evaluation of CLIP is identical to the setting of GUIClip.
Their only difference is the parameters' weights of the two models.

\item \textit{OCR + BGE}.
Rather than encoding the screenshot image, this method focuses on the textual information presented on the screenshots.
As described on the Section \ref{sec:rq1-baseline}, the texts displayed on screenshots are extracted with PaddleOCR \cite{Li:PPOCRv3MoreAttempts:2022}, then concatenated into sentence and embedded by BGE \cite{Xiao:CPackPackagedResources:2023}.
\end{itemize}

\subsubsection{Dataset}
We used the Enrico dataset to evaluate the model's performance for GUI classification.
Leiva et al.~\cite{Leiva:EnricoDatasetTopic:2020} used a random sample of 10k screenshots from Rico dataset to create Enrico (shorthand of Enhanced Rico): a human annotated dataset comprising 1460 screenshots and 20 design topics, including camera, chat, media player, etc. 
In the linear-probe setting, the Enrico dataset was divided into an 80:20 ratio to establish a training set and a test set.
For the zero-shot setting, all available data was employed solely for testing.

\subsubsection{Training Details of Linear-probe Classification}
For the training and evaluation of linear-probe classifier, we split the dataset into training and test set in a ratio of 80:20 in a stratified manner.
During training, the image encoder is frozen, the output embedding is served as input of the final linear layer.
We trained the model for 100 epochs using AdamW optimiser with an initial learning rate $1e^{-5}$, batch size 128.
We conducted a 10-fold cross-validation, which involved randomly splitting the training and testing sets ten times, and subsequently computing the average performance spanning across these runs.

\subsubsection{Evaluation Metric}
The evaluation of the classifiers is performed based on precision, recall, and the F1 score. 
Given the relatively small size of the Enrico dataset, especially when compared to the 20 available classification labels, we did not calculate the precision, recall, and F1 score for each individual label. 
In contrast, we calculate their weighted average.

\subsection{Exp4: Evaluation of GUIClip for Sketch-to-GUI Retrieval}
\label{sec:sketch-to-gui}
Instead of searching for a textual query, it is also possible to perform the GUI search using a UI sketch.
Based on GUIClip, we built a Dual GUIClip ViT model for sketch-to-GUI retrieval.
An evaluation on the Swire dataset \cite{Huang:SwireSketchbasedUser:2019} shows that our model outperform Swire and CLIP on sketch-to-GUI retrieval, with a recall@10 of 0.685.

\begin{figure}
    \centering
    \centerline{\includegraphics[width=1\textwidth]{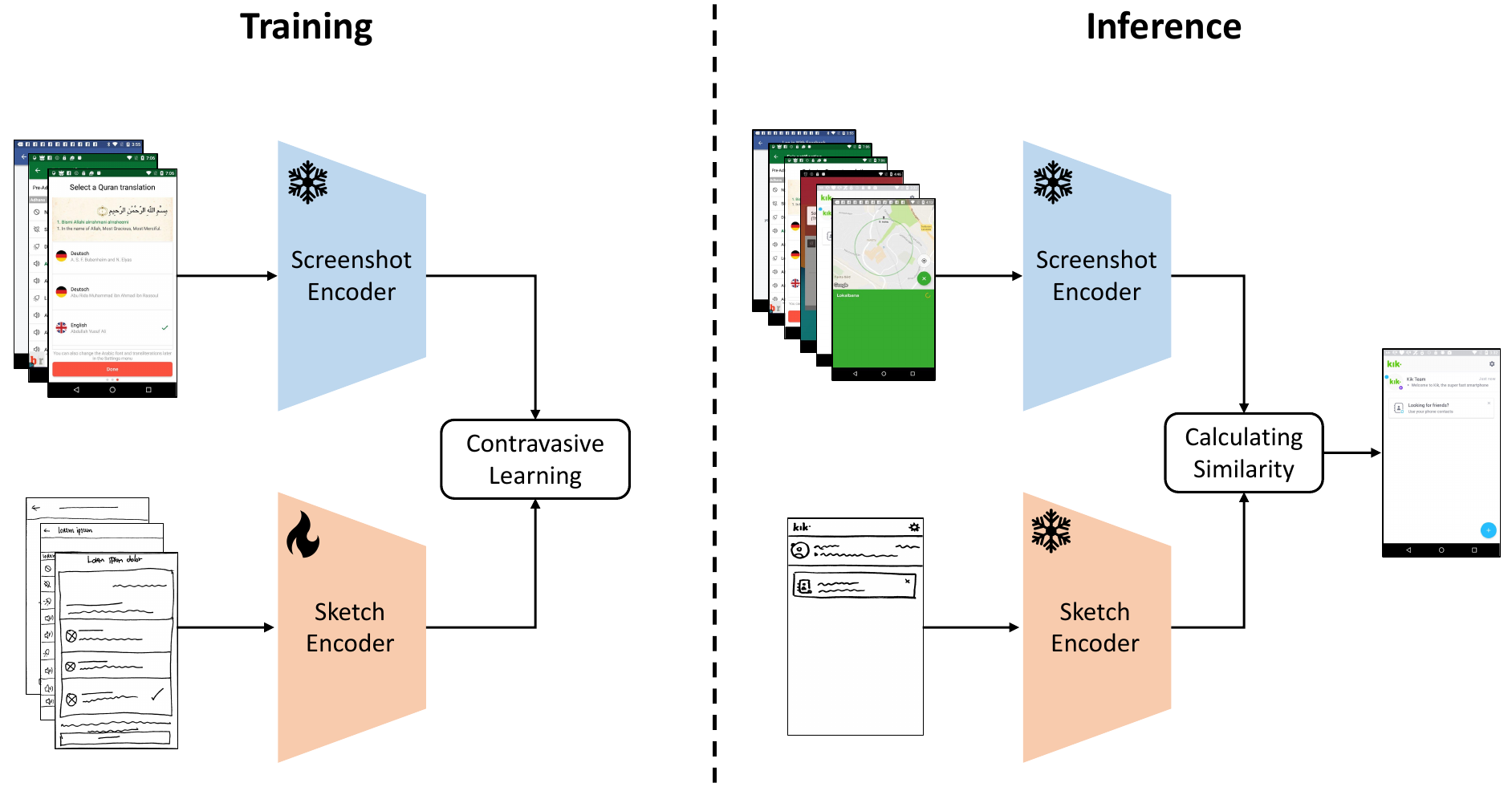}}
    \caption{Overview of the training and inference of Dual GUIClip ViT model for sketch-to-GUI retrieval.}
    \Description{Overview of the training and inference of Dual GUIClip ViT model for sketch-to-GUI retrieval.}
    \label{fig:GUIClip-sketch}
\end{figure}

\subsubsection{Model Architecture}
In order to adapt GUIClip for sketch-to-GUI retrieval, we employed its image encoder. 
The left side of Figure \ref{fig:GUIClip-sketch} illustrates the architecture of the model, which encompasses a screenshot encoder and a sketch encoder. 
Both of these components are initialised with the image encoder of GUIClip. 
During the training phase, the weights of the screenshot encoder are held constant, whilst training is solely conducted on the sketch encoder.
During the inference phase, which is illustrated at the right side of the Figure \ref{fig:GUIClip-sketch}, both encoders are frozen. 
The embedding of the query sketch is compared with the embeddings of each screenshot. 
The top-k screenshots whose embedding is the most similar with the sketch embedding are retrieved.

\subsubsection{Dataset}
The dataset employed in this study was procured from Swire \cite{Huang:SwireSketchbasedUser:2019}, which encompasses 3802 sketches corresponding to 2201 screenshots from 167 apps within the Rico dataset. 
This dataset was compiled by four designers, who contributed to sketching 505, 1017, 1272, and 1008 screenshots respectively. 
However, it was discovered that certain images from this dataset were not present within the Rico dataset. 
Hence after appropriate cleaning, a refined dataset featuring a total of 3551 sketches was obtained. 
The split between the training and test sets within Swire was not explicitly delineated in their work. 
Consequently, for the purposes of our study, we adopted 486 sketches from one designer as the test set, while the sketches from three other designers constituted the training set.
This ensured that the model learns to generalise across sketches developed by various designers.

\subsubsection{Baselines}
We compared GUIClip's performance in sketch-to-GUI retrieval with the following baselines:
\begin{itemize}
\item \textit{Swire}. Swire establishes itself as the pioneer sketch-to-GUI retrieval model \cite{Huang:SwireSketchbasedUser:2019}. 
This model comprises two distinct VGG-A networks \cite{Simonyan:VeryDeepConvolutional:2015} that independently compute the embeddings for screenshots and sketches. 
The retrieval process is subsequently conducted by measuring the similarity between these embeddings. 
As the implementation of Swire is not publicly accessible and as we were unable to get it from the authors, we decided to recreate it according to the details in the original paper. 

\item \textit{Dual CLIP ViT}.
The architecture of the Dual CLIP ViT is identical to that of GUIClip, except for their respective parameters' weights. 
The weights of the image encoder of Dual CLIP ViT are derived directly from the CLIP model.
\end{itemize}

\subsubsection{Training Details of Fine-tuning}
Under zero-shot setting, the Dual GUIClip ViT and Dual CLIP ViT have not been trained.
While under fine-tuning setting, the Dual GUIClip ViT and Dual CLIP ViT models have been trained with contrastive learning on the training set for 10 epochs.
We used AdamW optimiser with an initial learning rate $1e^{-5}$, batch size 256.

For the training of Swire, we tried to use the same configurations as specified in the original paper, including a learning rate of 0.01 and a batch size of 32. 
However, we noticed a significant decline in performance with these settings, wherein the top-10 recall stood at a mere 0.02. 
To address this, we decided to use a batch size of 64 and a learning rate of $1e^{-5}$.
As the original article does not specify the number of training epochs employed, we explored different values, ranging from 10 to 100 epochs in increments of 10. 
Our experiments indicates that 100 epochs produce the best results.

\subsubsection{Evaluation Metric}
Similar to the evaluation of text-to-GUI retrieval presented on the Section \ref{sec:eval-metric-recall}, we use recall@1, recall@3, recall@5 and recall@10.

\section{Evaluation Results}\label{sec:eval-result}

\begin{table}[]
\centering
\caption{Text-to-GUI retrieval performance on test set (Experiment 1).}
\begin{tabular}{ll cccccc}
\toprule
Dataset & Model & Recall@1 & Recall@3 & Recall@5 & Recall@10 & Recall@50 & Recall@100 \\
\midrule
\multirow{3}{*}{} & (Chance) & 0.001 & 0.003 & 0.005 & 0.010 & 0.050 & 0.100 \\
\midrule
\multirow{3}{*}{SCapRepo} 
& OCR + BGE & 0.164 & 0.291 & 0.355 & 0.441 & 0.624 & 0.705 \\
& CLIP     & 0.127 & 0.242 & 0.299 & 0.392 & 0.627 & 0.723 \\
& GUIClip-CS & 0.066 & 0.132 & 0.164 & 0.244 & 0.479 & 0.603 \\
& GUIClip   & \textbf{0.377} & \textbf{0.526} & \textbf{0.593} & \textbf{0.687} & \textbf{0.857} & \textbf{0.908} \\
\midrule
\multirow{3}{*}{Screen2Words} 
& OCR + BGE & 0.130 & 0.229 & 0.287 & 0.361 & 0.559 & 0.650 \\
& CLIP     & 0.097 & 0.171 & 0.220 & 0.298 & 0.508 & 0.615 \\
& GUIClip-CS & 0.195 & 0.349 & 0.429 & 0.532 & 0.800 & 0.887 \\
& GUIClip   & \textbf{0.216} & \textbf{0.371} & \textbf{0.449} & \textbf{0.566} & \textbf{0.825} & \textbf{0.903} \\
\midrule
\multirow{3}{*}{Clarity} 
& OCR + BGE & \textbf{0.152} & \textbf{0.242} & \textbf{0.281} & \textbf{0.338} & 0.500 & 0.585 \\
& CLIP     & 0.080 & 0.138 & 0.166 & 0.219 & 0.373 & 0.464 \\
& GUIClip-CS & 0.073 & 0.146 & 0.198 & 0.278 & 0.515 & 0.648 \\
& GUIClip   & 0.075 & 0.150 & 0.207 & 0.287 & \textbf{0.533} & \textbf{0.667} \\
\bottomrule
\end{tabular}
\label{tab:t2i-retrieval}
\end{table}

\subsection{RQ1: Performance of GUIClip in Text-to-GUI Retrieval}\label{sec:result-rq1}
Table \ref{tab:t2i-retrieval} shows the evaluation results for the text-to-GUI retrieval task using GUIClip (last row) and the baseline models on three distinct datasets. 
Compared to the retrieval by chance, which has a recall@10 of 0.01, the CLIP model proves to be more effective in text-to-GUI retrieval, with a recall@10 exceeding 0.2 across the three datasets. 
This implies that, given a caption, the CLIP model has a 20 percent probability that the corresponding screenshot lies within the top 10 retrieved screenshots.
Nevertheless, the GUIClip model consistently outperforms the CLIP model in almost all metrics and for the three datasets.
This indicates that our fine-tuned CLIP model, GUIClip, is more adept at bridging the semantic gap between the caption and the screenshot.

GUIClip-CS were trained merely on the Screen2Words and Clarity datasets. 
Notably, GUIClip delivers better performance than GUIClip-CS on the Screen2Words and Clarity datasets, registering a performance improvement ranging between 0.1 to 0.3 across various metrics. 
This performance distinction is further accentuated in the SCapRepo dataset, where GUIClip-CS exhibits an inferior performance, even when compared to the CLIP model. 
This outcome is hypothesised to derive from caption style variances. 
Specifically, the captions associated with the Screen2Words and Clarity datasets are significantly lengthy and detailed, which may possibly impede GUIClip-CS's capacity to generalise effectively on the SCapRepo dataset. 
On a holistic level, the results of GUIClip suggest that fine-tuning on our SCapRepo dataset contributes substantially to the performance of CLIP models for the text-to-GUI retrieval task. 
This dataset enhances the performance of the CLIP model not only on the SCapRepo dataset, but also leads to performance gains on other screenshot-caption datasets.

As for the text-only approach, when analysing the SCapRepo and Screen2Words datasets, it becomes evident that GUIClip surpasses the performance of OCR+BGE.
Interestingly, when the comparison is carried out on the Clarity datasets, GUIClip exhibits a deterioration in comparison to text-only approach for recall@1, Recall@3, recall@5 and recall@10. 
This discrepancy can presumably be attributed to the variability in caption styles.
The majority of captions in the SCapRepo dataset are app features such as ``Track your health trends'' or ``Custom colour themes''. 
And the captions found within the Screen2Words dataset generally encompass high-level descriptions of the screenshot, such as ``screen showing settings options'' and a ``display of news articles in a news app''.
In contrast, the Clarity dataset captions offer substantially more details about the screenshots, like ``at the bottom left there is a skip button for users to skip the introduction'', and ``in the middle of the screen a popup is displayed with a label called set date''. 
These detailed descriptions can easily be matched with the text displayed on the screenshots, thereby improving the performance of text-only approaches.
Nevertheless, GUIClip continues to surpass OCR+BGE in both recall@50 and recall@100.

Overall, our findings underscore that GUIClip outperforms the baseline models in text-to-GUI retrieval, particularly when the goal is to search for descriptions of an overall functionality and features implemented in the screens rather than verbose, documentation-like descriptions of the screen's UI elements.

\subsection{RQ2: Relevance of Search Results}\label{sec:result-rq2}

\begin{table}[]
\centering
\caption{Relevance of the retrieved screenshots by three GUI search engines (Experiment 2 / Manual assessments).}
\begin{tabular}{l ccccccccc}
\toprule
Search Engine & MRR & \multicolumn{4}{c}{P@k} & \multicolumn{4}{c}{HIT@k} \\
\cmidrule(lr){3-6}\cmidrule(lr){7-10}
 & & P@1 & P@3 & P@5 & P@10 & HIT@1 & HIT@3 & HIT@5 & HIT@10 \\
\midrule
RaWi & 0.410 & 0.291 & 0.264 & 0.254 & 0.214 & 0.291 & 0.487 & 0.594 & 0.701 \\
GUing-CS & 0.255 & 0.134 & 0.138 & 0.139 & 0.152 & 0.134 & 0.291 & 0.401 & 0.615 \\
GUing & \textbf{0.510} & \textbf{0.372} & \textbf{0.352} & \textbf{0.339} & \textbf{0.343} & \textbf{0.372} & \textbf{0.666} & \textbf{0.773} & \textbf{0.914} \\
\bottomrule
\end{tabular}
\label{tab:search-engine-retrieval}
\end{table}

Table \ref{tab:search-engine-retrieval} shows the results of the manual evaluation for GUing and the baseline search engines. 
The results show that GUing consistently outperforms RaWi across all evaluation metrics. 
P@1 and HIT@1 show that there is 0.372 probability that GUing will return a relevant screenshot as the first result, a considerable improvement over RaWi's rate of about 0.291. 
The P@10 value, indicative of the proportion of relevance results of the top-10 results returned by GUing, stands at 0.343, compared with RaWi's rate of 0.214. 
The HIT@10 evaluation metric shows that, in more than 91\% of cases, GUing can return at least one pertinent screenshot among the top-10 results, far exceeding the percentage for RaWi, which is recorded at 70\%.
This is particularly interesting for GUI ideation tasks, as it seems reasonable for a designer to check 10 screen suggestions for identifying at least one relevant screen (similar to screening the first result page of a google search). 
Finally, the MRR metric reveals that the ranking of the first relevant result returned by GUing is superior to that of RaWi.

The results also show that GUing significantly outperforms GUing-CS too.
This can be quantified by notable gaps in the following metrics: a gap exceeding 0.2 for P@k, approximately 0.3 for HIT@k, and 0.25 in terms of MRR.
The architectures of the search engines GUing and GUing-CS are identical, except the training set for the vision-language models and the repository used for searches. 
GUing-CS does not incorporate our SCapRepo and ScreenRepo datasets for its model training and search operation. 
The large gap between GUing and GUing-CS demonstrates the  importance of our SCapRepo and ScreenRepo datasets.

\begin{figure}[htbp]
    \centering
    \includegraphics[width=1\textwidth]{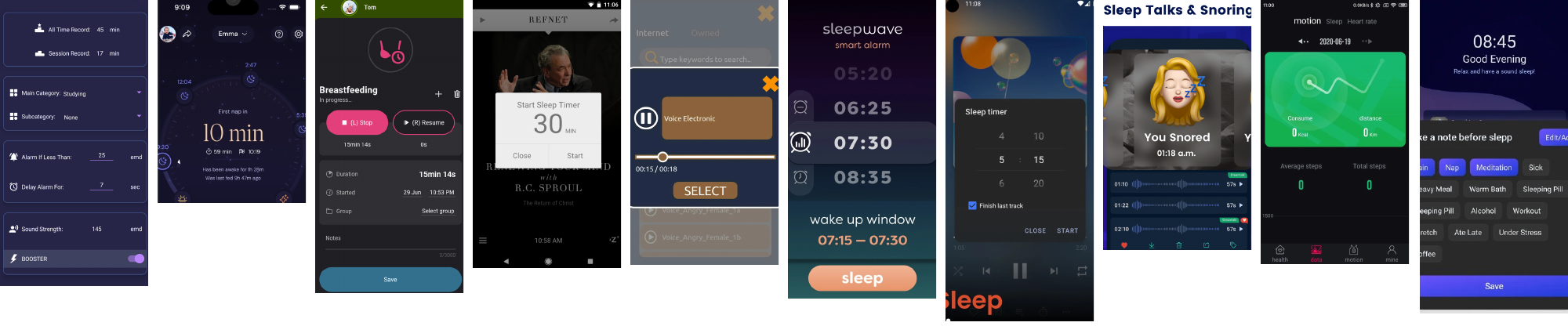}
    \caption{Top-10 screenshots returned by GUing for the query ``sleep tracking''.}\label{fig:guing-sample}
    \Description{Top-10 screenshots returned by GUing for the query ``sleep tracking''.}
    \bigskip
    \centering
    \includegraphics[width=1\textwidth]{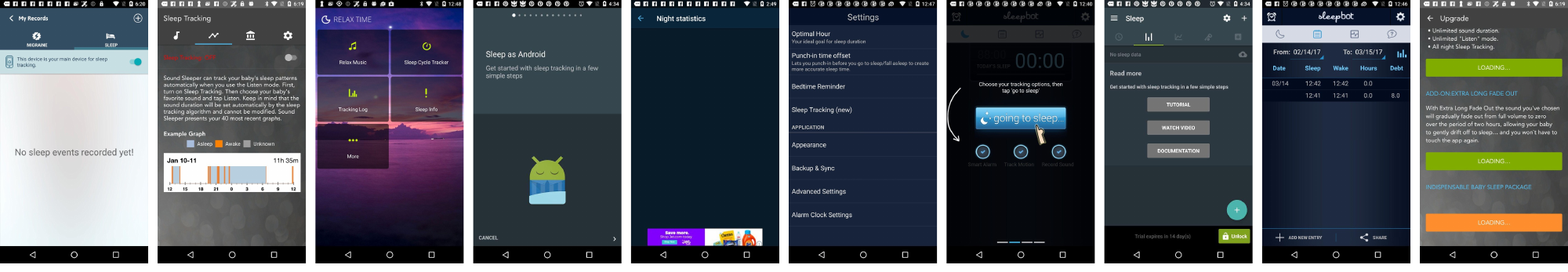}
    \caption{Top-10 screenshots returned by RaWi for the query ``sleep tracking''.}\label{fig:rawi-sample}
    \Description{Top-10 screenshots returned by RaWi for the query ``sleep tracking''.}
\end{figure}

To understand the difference between the screenshots obtained using GUing and Rawi, we carried out additional analyses on each search engine with a few queries.
For example, the screenshots returned by GUing and RaWi for the query "sleeping track" are shown on Figure \ref{fig:guing-sample} and \ref{fig:rawi-sample}.
Upon examination, it became apparent that there are considerable stylistic differences between the screenshots returned by the two search engines, mainly due to three factors:

\begin{itemize}
\item The shape of the screenshots returned by GUing is less neat.
A large portion of screenshots within the ScreenRepo dataset are derived from the \textit{surrounded screenshots}, in which the screenshot area varies in shape across diverse images.
However, these screenshots are still useful to inspire GUI design.

\item GUing returns a wider variety of screenshots than RaWi.
As shown on Figure \ref{fig:rawi-sample}, the presence of empty or settings pages among the screenshots returned by RaWi is thought to be an inherent limitation of the text-only retrieval approach, as they barely offer any contribution to sparking creativity in GUI design.
Text-only retrieval approaches compare the similarity of the query with text or metadata found on each screenshot.
As setting pages typically contain much text, they will likely be matched to many queries.
Empty pages are likely a result of the Rico data collection method, which involves the GUI exploration of apps.
When performing crowdsourced or automated exploration, the app often lack  initial data, such as recorded sleep events.

\item The modernity of the screenshots returned by GUing vastly exceeds those resulting from RaWi. 
This discrepancy can be attributed to the collection periods of the underlying datasets. 
The ScreenRepo dataset was collected recently, whereas the Rico dataset was created in 2017.
ScreenRepo can easily be updated by processing introduction images of new apps when added to the Google Play store.
\end{itemize}

Overall, our findings demonstrate that GUing consistently outperforms the baseline approaches in text-to-GUI retrieval task in term of relevance of the retrieved screens.

\subsection{RQ3: Performance of GUIClip in Other GUI-related Tasks}
\label{sec:result-rq3}

\subsubsection{GUI Classification}
As shown in Table \ref{tab:cls-enrico}, vision-language models, specifically CLIP and GUIClip, clearly outperform  OCR+BGE as expected. 
This reaffirms the assertion that mere analysis of text displayed on screenshots is insufficient for substantial GUI understanding, making the vision information crucial. 
Notably, GUIClip achieves superior results compared to CLIP, and records an F1 score that exceeds that of CLIP by 20 percent in both settings. 
This highlights the advantage of our GUIClip model for GUI classification tasks over the baseline models, given its enhanced knowledge about the GUI domain accumulated during the pre-training phase.
Particularly, the Linear-probe led to encouraging precision and recall of over 60\% (given the large number of classes in the classification task and the straight application of the model).

\begin{table}[htp!]
\caption{Classification accuracy on Enrico dataset.}
\begin{tabular}{l | ccc | ccc}
\toprule
& \multicolumn{3}{c}{Zero-shot} & \multicolumn{3}{c}{Linear-probe} \\
&Precision&Recall&F1 & Precision&Recall&F1\\
\midrule
OCR + BGE & 0.210 & 0.094 & 0.081 & 0.038 & 0.186 & 0.059 \\
CLIP    & 0.317 & 0.165 & 0.138 & 0.402 & 0.466 & 0.395 \\
GUIClip  & \textbf{0.415} & \textbf{0.347} & \textbf{0.334} & \textbf{0.613} & \textbf{0.640} & \textbf{0.600} \\
\bottomrule
\end{tabular}
\label{tab:cls-enrico}
\end{table}

The efficacy of vision-language models, as measured under the linear-probe setting, surpasses that determined under the zero-shot setting. 
This superior performance can be explained by the additional knowledge gained during the training phase from the Enrico dataset. 
However, a contrary pattern was found concerning OCR+BGE, which showed diminished performance following linear-probe training. 
We hypothesise that this is because the text-only approach is inadequate to extract visual information from the screenshots, which makes the training stage ineffective.

\subsubsection{Sketch-to-GUI Retrieval}
Table \ref{tab:ret-swire} presents the evaluation outcomes for GUIClip and the baseline models on the Swire dataset within both zero-shot and fine-tuning settings. 
Our observation revealed that even under zero-shot setting, the Dual GUIClip ViT has a recall@10 score of 0.224, which indicates its potential for sketch-to-GUI retrieval. 
In contrast, the Dual CLIP ViT was not as effective, as evidenced by its low recall@10 score of 0.053.

\begin{table}[htp!]
\caption{Retrieval accuracy on the test set of Swire dataset.}
\begin{tabular}{l | cccc | cccc}
\toprule
& \multicolumn{4}{c}{Zero-shot} & \multicolumn{4}{c}{Fine-tuned} \\
&Recall@1&Recall@3&Recall@5&Recall@10 &Recall@1&Recall@3&Recall@5&Recall@10\\
\midrule
Swire & - & - & - & - & 0.027 & 0.062 & 0.117 & 0.214 \\
Dual CLIP ViT & 0.019 & 0.029 & 0.037 & 0.053 & 0.177 & 0.296 & 0.422 & 0.533 \\
Dual GUIClip ViT & \textbf{0.062} & \textbf{0.117} & \textbf{0.156} & \textbf{0.224} & \textbf{0.368} & \textbf{0.580} & \textbf{0.685} & \textbf{0.772} \\
\bottomrule
\end{tabular}
\label{tab:ret-swire}
\end{table}

Fine-tuning can significantly improve the models' performance. 
After fine-tuning, the recall@10 score of the Dual CLIP ViT increased to 0.553, while the Dual GUIClip ViT exhibited a remarkable improvement, reaching 0.772. 
These results substantiate the superior efficacy of GUIClip in comparison to CLIP for sketch-to-GUI retrieval.

The performance of our Swire implementation, with a recall@1 of 0.027 and a recall@10 of 0.214, is not congruent with the metrics reported in the originating study by Huang et al.~\cite{Huang:SwireSketchbasedUser:2019}. 
The researchers reported a significantly higher recall values of 0.159 and 0.609, at ranks 1 and 10 respectively.
This discrepancy could potentially be attributed to two factors: 
1) Our evaluation used 486 pairings for testing, in contrast to the 276 pairings used in the original research.
The size of the test set can dramatically influence retrieval performance outcomes;
2) The implementation details could also lead to differing results. 
It is important to note, however, that the original Swire model's implementation is not available for replication.
As reported in Section \ref{sec:sketch-to-gui}, we tried our best to not disadvantage Swire in our evaluation as the original implementation was not accessible.
Nonetheless, even if we use the results from Huang et al.~\cite{Huang:SwireSketchbasedUser:2019}, our Dual GUIClip ViT model still demonstrates superior performance, with a score of 0.368 for recall@1 and 0.772 for recall@10.

\section{Related Work}

\subsection{Mobile GUI Retrieval}
GUI retrieval refers to the search of existing GUI designs (usually images) with queries of various format.

\textbf{By sketch image}:
UI sketches are often used at the early stage of the design. 
They can also be used as a query.
As detailed in Section \ref{sec:sketch-to-gui}, Huang et al.~introduced Swire \cite{Huang:SwireSketchbasedUser:2019}, which comprises two separate VGG-A networks \cite{Simonyan:VeryDeepConvolutional:2015} to compute the embeddings for screenshots and sketches, respectively. 
The retrieval process is subsequently performed by measuring the similarity of these embeddings.
Unlike Swire that treats a sketch image as a unit, Mohian et al.~\cite{Mohian:PSDoodleFastApp:2022,Mohian:SearchingMobileApp:2023} performed object detection on the sketches to identify their UI components, which are then used to find the screenshots in Rico dataset with the best matching types and locations of the UI components.

\textbf{By wireframe}:
A wireframe is an image that represents the skeletal layout of a screenshot.
Given a wireframe image, one can retrieve the corresponding screenshots. 
Deka et al.~\cite{Deka:RicoMobileApp:2017} and Liu et al.~\cite{Liu:LearningDesignSemantics:2018} trained an autoencoder \cite{Bengio:LearningDeepArchitectures:2009} for UI layout similarity, which supports query-by-example search over UIs.
Chen et al.~\cite{Chen:WireframebasedUIDesign:2020} performed wireframe-based GUI retrieval by encoding the visual semantics of UI designs using a large database of UI design wireframes.

\textbf{By screenshot image}:
Using a screenshot from an app, it is possible to retrieve similar screenshots from a GUI repository.
Screen2vec \cite{Li:Screen2vecSemanticEmbedding:2021} extracts the UI components, layout, and app description to generate a screen embedding. 
Querying similar screenshots is then done by comparing the similarity of screen embeddings.
VINS \cite{Bunian:VINSVisualSearch:2021} creates screen embeddings by combining both image embeddings and UI components embeddings.
Instead of querying screenshots, GUIFetch \cite{Behrang:GUIfetchSupportingApp:2018} searches for apps in public repositories that include similar screens and transitions between them.

\textbf{By text}:
Using textual queries for GUI retrieval is practical as often only a textual description or keywords are available at the search time. 
GUIGLE \cite{Bernal-Cardenas:GuigleGUISearch:2019} has indexed a collection of GUIs along with their metadata from the ReDraw dataset \cite{Moran:MachineLearningBasedPrototyping:2020}.
GUIGLE leverages various information sources such as text displayed on the screen, names of UI components, and app name to retrieve relevant screens.
On the other hand, as presented in Section \ref{sec:exp2}, RaWi \cite{Kolthoff:DatadrivenPrototypingNaturallanguagebased:2023} uses a BERT-based \cite{Devlin:BERTPretrainingDeep:2019} ranking model to retrieve GUIs from the Rico dataset \cite{Deka:RicoMobileApp:2017}. 
This retrieval process involves matching the GUI text (text displayed on the screen, GUI activity name, and GUI component identifier) with the textual query in order to identify the most relevant images.
Unlike the preceding studies that focus on retrieving entire screenshot, Gallery D.C.~\cite{Chen:GalleryDesignSearch:2019,Feng:GalleryAutocreatedGUI:2022} allows users to search UI components, like a button or a checkbox.
It provides a GUI gallery that allows users to search UI components with different filters, including component type, size, colour, app category, and displayed text.

All GUI retrieval approaches suggested so far either use the visual information or the textual information available about the the mobile GUIs for matching the query with corresponding data in the repository. 
In contract, our approach combines both modalities by training a text-vision model based on screen-caption pairs. 
This bridges between the visual perception and the natural language and boosts the retrieval performance as our results show.

\subsection{Vision Language Models for Mobile GUI}
Vision-language models combines both the vision and language modalities. 
They are trained on image-text pairs and can be applied to a range of tasks \cite{Du:SurveyVisionLanguagePreTrained:2022,Zhang:VisionLanguageModelsVision:2024}.
Until this work, the main use case of vision-language models in the mobile GUI domain is the GUI captioning.
This refers to the task of generating a high-level summary of a screenshot to describe its contents and functionalities.
The Screen2Words model \cite{Wang:Screen2WordsAutomaticMobile:2021}, built on the foundations of the Transformer Encoder-Decoder architecture \cite{Vaswani:AttentionAllYou:2017}, uses multimodal data, inclusive of the screenshot image, view hierarchy, and app description, to yield the screen caption.
Contrarily, Spotlight, a vision-only approach employed for multiple tasks including widget captioning and screen captioning, possess architecture rooted in Vision Transformer \cite{Dosovitskiy:ImageWorth16X16:2021} and the T5 model \cite{Raffel:ExploringLimitsTransfer:2020}.

While some studies pertaining to GUI-related tasks may appear to employ vision-language models, they do not in reality.
For instance, XUI \cite{Leiva:DescribingUIScreenshots:2023} model does not fully address the language modality. 
It generates the screen caption with template-based natural language generation (NLG) engine.
As discussed above, GUIGLE \cite{Bernal-Cardenas:GuigleGUISearch:2019} and RaWi \cite{Kolthoff:DatadrivenPrototypingNaturallanguagebased:2023} introduced text-based GUI retrieval approaches by using the textual content exhibited in the screenshots or the metadata, instead of employing the vision information for the search.
This may result in the omission of crucial vision information of the UI screen.
To the best of our knowledge, no existing research employs vision-language models for text-to-GUI retrieval.

Recently, two related vision-language models, UIClip \cite{Wu:UIClipDatadrivenModel:2024} and Ferret-UI \cite{You:FerretUIGroundedMobile:2024}, have been presented in parallel to our work.
UIClip, like our model GUIClip, is a fine-tuned version the CLIP model. 
However, their purposes differ significantly. 
While GUIClip is designed for GUI retrieval (as app vendors describe what is implemented on the corresponding screen), UIClip is trained to assess the design quality and visual relevance of a UI.
Ferret-UI, on the other hand, is a multimodal foundation model that goes beyond vision-language models by supporting more input formats, such as points, boxes, and scribbles on screenshots. 
This model is capable of performing various tasks (e.g., widget classification, icon recognition, OCR) with flexible input formats, as well as grounding tasks (e.g., finding widgets, icons, or text, and listing widgets on a screen).

\subsection{Mobile GUI Datasets}
Large-scale mobile GUI datasets comprise a vast collection of screenshots and serve as a valuable resource for GUI retrieval. 
Rico \cite{Deka:RicoMobileApp:2017,Liu:LearningDesignSemantics:2018} is one of the largest datasets in the literature.
Deka et al.~\cite{Deka:RicoMobileApp:2017} combined crowdsourcing and automation to mine design and interaction data from Android apps at runtime.
Their dataset exposes visual, textual, structural, and interactive design properties of more than 66k unique UI screens appx.~9.3k Android apps.
Several other datasets \cite{Leiva:EnricoDatasetTopic:2020,Sunkara:BetterSemanticUnderstanding:2022,Zaki:MASCDatasetDevelopment:2023} are built on Rico including the Screen2Words dataset \cite{Wang:Screen2WordsAutomaticMobile:2021} used in this paper.
As Rico was created seven years ago, some of its UIs might be outdated for contemporary apps as our evaluation suggests.

Chen et al.~\cite{Chen:WireframebasedUIDesign:2020} developed an extensive dataset that comprises almost 55k screenshots from 7,748 Android apps.
The screenshots were obtained through automated GUI exploration \cite{Chen:UIDesignImage:2018}. 
Moran et al.~\cite{Moran:MachineLearningBasedPrototyping:2020} collected the ReDraw dataset in a fully automated manner by mining and executing the top-250 Android apps in each category on Google Play, excluding games. 
This yielded a total of 14,382 unique screenshots and 191k labelled GUI components. 
Clarity \cite{Moran:EmpiricalInvestigationUse:2022} is an annotated subset of ReDraw consisting of 45,998 descriptions for 10,204 screenshots of popular apps. 
The descriptions were created by crowd workers using Amazon Mechanical Turk. 
Instead of focusing on entire screens, Li et al.~\cite{Li:WidgetCaptioningGenerating:2020} released a dataset containing appx.~162,859 phrases created by human workers for annotating 61k UI components across 21,750 unique screens.

Generally, more and more crowdsourced or automated GUI exploration methods are used to (semi)automatically explore UIs of iOS apps \cite{Wu:NeverendingLearningUser:2023} or Android apps \cite{Chen:StoryDroidAutomatedGeneration:2019,Chen:AutomaticallyDistillingStoryboard:2023,Zhang:SceneDrivenExplorationGUI:2023}. 
While certainly efficient and scalable, automated UI exploration rather targets UI testing than GUI design inspiration.  
GUI datasets gathered through crowdsourced or automated exploration may omit important app features since the access to certain UIs may necessitate app authorisation or initial configurations  \cite{Chen:AutomaticallyDistillingStoryboard:2023}.
In addition, these datasets are not curated by actual app designers and vendors. 
Recent studies showed that app vendors and app users (or the crowd) often use different vocabularies to describe app features \cite{Haering:AutomatingEvaluationEducation:2021,Haering:AutomaticallyMatchingBug:2021}. 
The dataset introduced in our work ScreenRepo and SCapRepo aims at closing this gap. 
While we also used automated techniques for the data collection and cleaning, the screen-caption pairs in SCapRep are curated by various app vendors, describing what a screen is supposed to do and what problem for the user does it solve.

\section{Discussion}
We discuss the implications of our work for research and practice together with potential threats to validity.

\subsection{Implications for Software Practitioners}
Our work has several potential implications for software practitioners, particularly for app designers and requirements engineers.
Our search engine primarily enables app designers and developers to get inspirations on how their screens can look like, for instance in an early project phase, when only a list of app features is available \cite{Wei:GettingInspirationFeature:2024}. 
Particularly in small development teams, which often lack resources and expertise for GUI design and usability engineering \cite{bornoe2013supporting}, this can provide a substantial support.
Despite the general trend with Generative AI for more code generation and automation, we think that it remains crucial in professional app development to have the designer (or developer) in the loop \cite{Wei:AIInspiredUIDesign:2024}. 
GUing is such a tool that uses AI to speed-up design while ensuring the crucial role of human creativity and oversight.

GUing can also serve as a rapid prototyping tool. 
It can, e.g., be used during requirements interviews and in workshops, to retrieve screenshots based on identified requirements, speeding up their refinement and validation. 
This by itself, increases stakeholder engagement and helps mitigate development risks. 
Additionally, retrieved screenshots serve as a visual aid for illustrating and documenting the requirements and reducing potential ambiguity \cite{Montgomery:EmpiricalResearchRequirements:2022}.

Given the significant impact of GUI on user experience and retention \cite{Chen:HowShouldImprove:2021}, and as GUI designs and user preferences evolve rapidly, it is essential for app designers to continuously observe and learn from trends and best practices. 
GUing leverages app introduction images curated by actual app vendors and experts. 
Our pipeline for image processing and model training allows for a seamless addition of new images to the GUing repository, ensuring a high quality and up-to-dateness. 
This helps designers learn the latest UI design trends.
When augmented with download, rating, and feedback trends, retrieved screens can help not only get inspiration but also estimate user trends \cite{Martens:ReleaseEarlyRelease:2019}.

\subsection{Implication for Software Researchers}
There has been extensive research on app store mining over the last decade, with the majority of studies focusing on app reviews and descriptions \cite{Al-Subaihin:AppStoreEffects:2021,Martin:SurveyAppStore:2017,Maalej:AutomatedProcessingUser:2024}.
Less work have delved into the analysis of other developer information shared in the stores, such as app introduction images \cite{Chen:GalleryDesignSearch:2019,Feng:GalleryAutocreatedGUI:2022}. 
Our work highlights the potential of this information, particularly as it is curated, usually exhibits a high quality (as created by professionals), and captures various modalities (image, text, videos, binaries, etc.).
We think that researchers should focus more on investigating this information and combining it with crowd-data for building multimodal foundation models for software engineering.

From the app introduction images, we developed and shared the SCapRepo dataset and subsequently the vision-language model GUIClip. 
The evaluation results confirms the power for these models to learn and combine modalities, not only for screenshot retrieval but also at least for GUI classification and sketch-to-GUI retrieval. 
We think that GUIClip and similar foundation models tuned on software design data have the potential to be adapted for other GUI-related tasks, including GUI captioning, GUI generation, or GUI testing. 
As a foundational model, CLIP model itself is the basis to numerous models \cite{Zhang:VisionLanguageModelsVision:2024}, such as the image captioning model ClipCap \cite{Mokady:ClipCapCLIPPrefix:2021} and the image generation model Stable Diffusion \cite{Rombach:HighResolutionImageSynthesis:2022}. 
By replacing the CLIP model with tuned, domain specific models such as  GUIClip, it is possible to enhance these derivative models' capabilities for specific tasks.
Researchers should investigate this in details, not only with regard to model performance (or accuracy) but also with regard to quality of output, risk of prediction faults, as well as design and engineering workflows to be adapted and integrated \cite{Wei:AIInspiredUIDesign:2024,Maalej:TaskFirstContextFirstTool:2009}.

\subsection{Threats to Validity}
This section discusses potential threats to the validity of our work.

\subsubsection{Internal and Construct Validly}
As for every study that includes manual evaluation, the process of evaluating the usefulness of the search results may have included inherent biases. 
Particularly, the evaluation outcomes may have been influenced by the evaluators' preferences towards specific search engines. 
In an attempt to mitigate this potential observer bias, we incorporated a number of steps. 
We created an evaluation tool which mixes and shuffles the resulting screenshots originating from the three distinct search engines. 
As a result, the evaluator cannot discern the origin of a specific screenshot, thereby curbing potential bias among diverse search engines.
Moreover, the evaluation process was carried out by four evaluators, each with a minimum of five years' experience in software development.
They conducted a careful evaluation based on a evaluation guide (e.g.~stating what is a relevant GUI and what is not) and using a uniform evaluation tool.
Finally, each query was independently assessed by at least two evaluators to minimise potential mistakes.

\subsubsection{External Validity}
We discuss the threats to external validity for the evaluation of GUing and GUIClip.

\textit{GUing: Query Selection for Search Engine Evaluation.}
The selection of queries is crucial for the evaluation, since the search performance may vary depending on the query.
For example, a GUI search engine that exhibits satisfactory results for the health and fitness domain may not necessarily produce the same outcome for the finance or education domains. 
It is thus imperative to consider varying domains and apps in the evaluation. 
The study of Kolthoff et al.~\cite{Kolthoff:DatadrivenPrototypingNaturallanguagebased:2023} provides a gold standard of queries and corresponding GUIs.
However, this dataset has already been exhausted for the training of RaWi, which makes it unsuitable for our comparative study. 
Given the lack of alternative query datasets, we searched for articles published by companies specialised in mobile app development. 
This decision could potentially introduce a selection bias.
The app features extracted from the articles span ten distinct domains \cite{138AppFeatures}, plus innovative features from multiple additional domains \cite{50AppFeatures}. 
Therefore, we think that these sets are representative enough for the purpose of our evaluation.
Nevertheless, reproducing our study with additional queries, scenarios, and evaluators would certainly further strengthen the generalisability of its results.

\textit{GUIClip: Bias of Datasets for Model Evaluation.}
Enrico contains 1460 screenshots for 20 categories, while Swire contains 3551 sketch-screenshot pairs.
Compared with the large scale of SCapRepo, these datasets used to evaluate GUIClip for other GUI-related tasks are relatively small. 
This may introduce potential bias into evaluation.
To address this, we manually reviewed these two datasets and found that they encompass diverse domains. 
This suggests that GUIClip is effective for handling various GUIs across multiple domains. 
However, to strengthen the external validity of the findings, further evaluation on additional datasets is necessary, ideally with a broader range of domains, apps, and designs.

\section{Conclusion and Future Work}
In this paper, we presented GUing, a GUI search engine, built upon GUIClip, a novel vision-language foundation model for the domain of mobile apps. 
The GUIClip model was trained using SCapRepo, a comprehensive dataset comprised of 135k screenshot-caption pairs created by actual app experts. 
In addition, we presented ScreenRepo, a dataset that serves as a large repository for GUing. 
SCapRepo and ScreenRepo were created through an automated pipeline, enabling an easy update, e.g.~when new screen designs emerge. 

Our evaluation, including a benchmarking on various datasets and a manual assessment of search results, indicates that GUing outperforms state-of-the-art approaches for GUI retrieval.
Moreover, the evaluation highlights the critical role that the SCapRepo and ScreenRepo datasets play in achieving the top performance of our vision-language model: not only for GUI retrieval but also for GUI classification and sketch-to-GUI retrieval in the mobile app domain. 
From here, there are several follow-up future directions: 

\begin{itemize}
\item \textit{Increasing and diversifying the screenshot dataset.}
The performance of a vision-language model is correlated with the size of its training data. 
According to Statista \cite{Statista-Google-Play}, Google Play includes approximately 2.43 million apps as of 2023. 
Our training dataset includes merely 117k apps. 
Mining additional app introduction images will thus likely improve the performance of GUing and relevance of the results. 
Moreover, other platforms like Apple iOS, Samsung Galaxy Watch, or Microsoft HoloLens also have stores with similar app introduction images. 
Learning and searching those screenshots can improve the engine and inspire the design of corresponding apps. 

\item \textit{Hybrid search methods.}
The inherent limitations of natural language can pose difficulties in precisely describing the desired GUIs through text. 
To enhance the retrieval process, it might be beneficial to support a more diverse range of search methods and data. 
For instance, sketch-to-GUI retrieval would enable users to find GUI designs (or related follow up screens). 
Additionally, GUI-to-GUI retrieval would enable users to retrieve other GUIs that exhibit a resemblance (or other dependencies) to the original interfaces. 

\item \textit{Evaluations with practitioners and integration into the development workflows.}
Although our empirical evaluation show the efficacy of GUing, the reality of software development can be more complex and nuanced.
Thus, additional empirical studies with app developers, UI designers, and requirements engineers would likely lead to comprehensive feedback on real usage, scalability, further search parameters, as well as how GUing and GUIClip can be ideally integrated into existing app development workflows and tools \cite{Maalej:TaskFirstContextFirstTool:2009}. 
\end{itemize}

\bibliographystyle{ACM-Reference-Format}
\bibliography{ref}

\end{document}